\documentclass[a4paper,fleqn,usenatbib]{mnras}
\usepackage{newtxtext,newtxmath}
\usepackage[T1]{fontenc}
\usepackage{ae,aecompl}
\usepackage{float}
\usepackage{graphicx}   
\usepackage{amsmath}	
\usepackage{amssymb}	

\usepackage{gensymb}    
\usepackage{booktabs}   
\usepackage{url}         
\usepackage{trimclip}
\usepackage{epstopdf}
\usepackage{color,colortbl}
\usepackage{threeparttable}
\title[QZ\,Lib: a period bouncer]
      {The cataclysmic variable QZ\,Lib: a period bouncer}
\author[A.~F. Pala, L. Schmidtobreick, C. Tappert, B.~T. G\"ansicke, A. Mehner]
      {A.~F.~Pala$^{1}$\thanks{A.F.Pala@warwick.ac.uk}, 
       L.~Schmidtobreick$^{2}$\thanks{lschmidt@eso.org},
       C.~Tappert$^{3}$, 
       B.~T.~G\"ansicke$^{1}$ 
       and A.~Mehner$^{2}$\\
       $^{1}$Department of Physics, University of Warwick, Coventry, CV4 7AL, UK\\
       $^{2}$European Southern Observatory, Casilla 19001, Santiago 19, Chile\\
       $^{3}$Institute of Physics and Astronomy, Universidad de Valpara{\'i}so, Av. Gran Bretana 1111, Valparaiso, Chile}

\date{Accepted 2018 September 03. Received 2018 September 03; in original form 2018 January 17.}
\pubyear{2009}

\begin{document}
\pagerange{\pageref{firstpage}--\pageref{lastpage}}
\maketitle
\label{firstpage}

\begin{abstract}
While highly evolved cataclysmic variables (CVs) with brown dwarf donors, often called ``period bouncers'', are predicted to make up $\simeq40-70$\% of the Galactic CV population, only a handful of such systems are currently known. The identification and characterization of additional period bouncers is therefore important to probe this poorly understood phase of CV evolution. 
We investigate the evolution of the CV QZ\,Lib following its 2004 super--outburst using multi--epoch spectroscopy. From time--resolved spectroscopic observations we measure the orbital period of the system, $P_\mathrm{orb}= 0.06436(20)$\,d, which, combined with the superhump period $P_\mathrm{SH}= 0.064602(24)$\,d, yields the system mass ratio, $q = 0.040(9)$.
From the analysis of the spectral energy distribution we determine the structure of the accretion disc and the white dwarf effective temperature, $T_\mathrm{eff} = 10\,500 \pm 1500\,\mathrm{K}$. We also derive an upper limit on the effective temperature of the secondary, $T_\mathrm{eff} \lesssim 1700\,\mathrm{K}$, corresponding to a brown dwarf of T spectral type. The low temperature of the white dwarf, the small mass ratio and the fact that the donor is not dominating the near--infrared emission are all clues of a post bounce system.
Although it is possible that QZ\,Lib could have formed as a white dwarf plus a brown dwarf binary, binary population synthesis studies clearly suggest this scenario to be less likely than a period bouncer detection and we conclude that QZ\,Lib is a CV that has already evolved through the period minimum.
\end{abstract}

\begin{keywords}
stars: dwarf novae, binaries: close, brown dwarfs -- stars: individual: QZ\,Lib
\end{keywords}

\section{Introduction}
Cataclysmic variables (CVs) are close interacting binaries composed by a white dwarf (the primary) and a low--mass star (the donor or secondary). The secondary star fills its Roche--lobe and is therefore losing mass through the inner Lagrangian point. In the absence of strong magnetic fields, this matter is then accreted onto the white dwarf via an accretion disc (see \citealt{warner95-1} for a comprehensive review).

For the mass transfer process to be stable, the system needs to lose angular momentum in order to continuously shrink and keep the secondary in touch with its Roche--lobe. Therefore, during their life, CVs evolve from long to short orbital periods ($P_\mathrm{orb}$) while angular momentum is removed from the system by two mechanisms: magnetic braking and gravitational wave radiation \citep{rappaportetal83-1,paczynski+sienkiewicz83-1,spruit+ritter83-1}. The evolution proceeds towards shorter periods until the system reaches the ``period minimum'', which theory predicts to be $P_\mathrm{orb} \simeq 70\,\mathrm{min}$ \citep{Goliasch-Nelson2015,Kalomeni2016}. At this stage the timescale at which the secondary star loses mass becomes much shorter than its thermal timescale and thus the secondary stops shrinking in response to the mass loss. Consequently, systems that have passed the period minimum evolve back towards longer orbital periods and, for this reason, are called ``period bouncers''. 

Although this standard model of CV evolution is widely accepted, a number of discrepancies between theory and observations suggest that our current understanding of CV evolution remains incomplete. One of the major disagreements is found near the period minimum. The time that a CV spends at a given orbital period (and thus its detection probability at that $P_\mathrm{orb}$) is inversely proportional to the rate at which $P_\mathrm{orb}$ varies. The evolution of CVs approaching the period minimum, and evolving back towards longer periods, is much slower than that of younger systems. Therefore an accumulation of systems is expected to be observed at the period minimum (the so--called ``period minimum spike'').
Thanks to large sample of faint systems detected by the Sloan Digital Sky Survey \citep{Szkody_etal_2002, Szkody_etal_2003, Szkody_etal_2004, Szkody_etal_2005, Szkody_etal_2006, Szkody_etal_2007, Szkody_etal_2009, Szkody_etal_2011}, \citet{Gaensickeetal2009} confirmed the existence of this predicted pile--up of CVs at the period minimum. However, they also demonstrate that the observed period minimum ($\simeq 83\,\mathrm{min}$) is actually longer than the theoretically predicted ($\simeq 70\,\mathrm{min}$). Second, $\simeq 40\%$ \citep{Goliasch-Nelson2015} to $\simeq 70\%$ \citep{kolb93-1,kniggeetal11-1} of the present Galactic CV population is expected to have evolved past the period minimum. However, despite there are now more than 1400 CVs with an orbital period determination \citep{ritter+kolb03-1}, only a handful of period bouncer \textit{candidates} have been so far identified (e.g. \citealt{pattersonetal05-2,Unda-Sanzana2008, Littlefair2006, Patterson2011, katoetal2015, katoetal2016, McAllister2017, Neustroev2017}). 
Therefore the identification of additional period bouncers and the determination of their physical parameters is important for a direct comparison between the current models of CV evolution and observations in a region of the parameter space where very few objects are known.

One of the best parameters to identify a period bouncer is the mass ratio $q = M_\mathrm{sec}/M_\mathrm{WD}$. During the system evolution, donor stars in CVs continuously lose mass and, since the mass distribution of CV white dwarf is comparatively narrow and centred at $\simeq 0.8\,\mathrm{M}_\odot$ \citep{Zorotovic}, the mass ratio reflects the evolutionary stage of the system \citep{pattersonetal05-2,kniggeetal11-1}. In particular, for systems with $P_\mathrm{orb} \simeq 90\,\mathrm{min}$ that have not yet evolved through the period minimum (pre--bounce CVs), the expected mass ratio is $q \simeq 0.13$ while period bouncers at the same orbital period are expected to have $q \simeq 0.03$ \citep{Knigge2006}. Given the bifurcation of the CV evolutionary track at the period minimum, this difference further increases for longer period CVs (see for example figure 6 from \citealt{howelletal01-1}). Therefore the mass ratio is a discriminant between pre and post--bounce systems and provides a powerful tool to successfully identify period bouncer candidates when a direct spectroscopic detection of the secondary is not possible \citep{Patterson2011,katoetal2015,katoetal2016}.

A measurement of the mass ratio can be obtained from superhumps, i.e. low--amplitude modulations observed in the light curve of short--period CVs during superoutbursts. These are sudden increases in the system brightness originating from the combination of a thermal instability in the disc and the tidal interaction of its outer edge with the secondary \citep{osaki96-1}. Below a critical mass ratio value ($q \simeq 0.3$, \citealt{whitehurst88-1}), owing to the strong secondary tidal torque, the disc becomes elliptical and starts precessing, giving rise to the superhumps.
Superhumps have periods ($P_\mathrm{SH}$) typically a few percent longer than the orbital one. \cite{pattersonetal05-3} and \cite{kato2013} calibrated an empirical relationship that allows to measure the system mass ratio from this period excess.

In February 2004, Pojmanski (VSNET--alert 7982) detected the outburst of a new eruptive star in Libra, QZ\,Lib (aka ASAS\,153616--0839.1). Light curves taken worldwide revealed that the object showed periodic variability with a probably increasing period \citep{Kiyota04-1}, which was interpreted as growing superhumps with a final period of $P_\mathrm{sh} = 0.06501(3)$\,d \citep{kato04-1}.
QZ\,Lib was then spectroscopically confirmed of being a dwarf nova into outburst by \citet{schmidtobreicketal04-1}.

QZ\,Lib was included by \citet{Patterson2011} in his list of period bouncer candidates (see his table~3 and table~5) owing to its estimated low mass ratio ($q = 0.035 \pm 0.020$, \citealt{pattersonetal05-2}). However, \citet{pattersonetal05-2} obtained this measurement from unpublished CBA (Center for Backyard Astrophysics) data which where characterised by low signal--to--noise ratio (SNR). Here, we present new phase--resolved photometric and spectroscopic observations of this system, from which we derive a more reliable mass ratio that, combined with the spectral energy distribution analysis, demonstrates the period bouncer nature of the system.

\begin{table*}
\caption{Summary of the observational details. The last column reports the time range spun by the individual dataset for a comparison with the orbital period of QZ\,Lib, 1.54\,h.}\label{obstab}
 \begin{tabular}{@{}llllllll@{}}
  \toprule
Date (UT) & Tel./Inst. & Grism/Slit/Filter & $R = \lambda/ \Delta \lambda$ & $N_{\rm Exp}$ & $t_{\rm Exp}$\,[s] & $\Delta t$\,[h] \\
\midrule
2004-03-15 & 3.6/EFOSC2 & Gr\,\#6, 1.0" &323 &3 & 600 & 0.53\\
2004-03-16 & 3.6/EFOSC2 & Gr\#10, 1.0" &915 &30 & 400 & 3.62\\
2004-03-17 & 3.6/EFOSC2 & Gr\#10, 1.0" & 915&15 & 300 & 1.40\\
2004-05-01 & 3.6/EFOSC2 & Gr\#6, 1.0" & 323 &3 & 600 & 0.53\\
2004-08-27 & 3.6/EFOSC2 & Gr\#6, 1.0" & 323 &3 & 900 & 0.78\\
2004-08-28 & 3.6/EFOSC2 & Gr\#10, 1.0" &915 &6 & 400 & 0.72\\
2004-08-29 & 3.6/EFOSC2 & Gr\#10, 1.0" & 915 &26 & 400 & 3.14\\
2004-08-30 & 3.6/EFOSC2 & Gr\#10, 1.0" & 915 &11 & 400 & 1.33\\
 \noalign{\smallskip}
2005-02-07 & 1.0--m SMARTS & $V$ & &12 & 300 & 1.08\\
2005-02-09 & 1.0--m SMARTS & $V$ & &19 & 300 & 1.71\\
2005-02-10 & 1.0--m SMARTS & $V$ & &21 & 300 & 1.90\\
2005-02-11 & 1.0--m SMARTS & $V$ & &17 & 300 & 1.53\\
2005-02-12 & 1.0--m SMARTS & $V$ & &20 & 300 & 1.80\\
2005-02-13 & 1.0--m SMARTS & $V$ & &22 & 300 & 1.99\\
 \noalign{\smallskip}
2005-02-13 & 3.6/EFOSC2 & $V$ & & 145 & 20 & 2.13\\
 \noalign{\smallskip}
2012-08-13 & VLT/X--shooter &UVB,VIS,NIR  1.0", 0.9", 0.9" & 5400, 8900, 5600 & 6,6,7 & 244,252,249 & 0.50\\
2012-08-15 & VLT/X--shooter &UVB,VIS,NIR  1.0",  0.9", 0.9" & 5400, 8900, 5600 &4,4,4 & 244,252,249 & 0.33\\
2013-04-21 & VLT/X--shooter &UVB,VIS,NIR 1.0",  0.9", 0.9" & 5400, 8900, 5600 &24,24,24 & 244,252,249 & 1.98\\
2015-05-12 & VLT/X--shooter &UVB,VIS,NIR 1.0",  0.9", 0.9" & 5400, 8900, 5600 &9,7,7 & 410,465,510 & 1.11\\
\bottomrule
\end{tabular}
\begin{tablenotes}
\item \textbf{Notes.} $\Delta t$ includes the overhead times: $\simeq 34\,$s for EFOSC2 imaging and spectroscopy, $\simeq 25\,$s for SMARTS imaging and $\simeq 45\,$s for X--shooter spectroscopy. \\
\end{tablenotes}
\end{table*}

\begin{figure}
\centering
\includegraphics[width=0.48\textwidth]{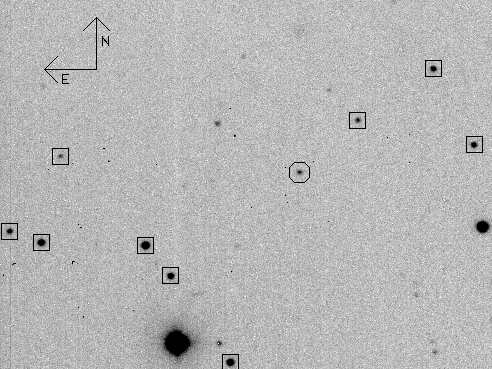}
\caption{SMARTS $V$--band finding chart of QZ\,Lib (circle, $\mathrm{RA} =15^{\mathrm{h}}\,36^{\mathrm{m}}\,16\fs\,0$, ${\mathrm{Dec.}} = -08\degree\,39\arcmin\,07\fs\,6$) and the nine reference stars used for the differential photometry (squares). The field of view is $4^\prime \times 3^\prime$, north is up, east to the left.}\label{chart}
\end{figure}

\section{Observation and data reduction}
\subsection{Spectroscopic observations}\label{subsec:spec_obs}
Spectroscopic observations of QZ\,Lib were obtained at several epochs in 2004 using EFOSC2 (ESO Faint Object Spectrograph and Camera 2) at the 3.6--meter telescope on La Silla Observatory, Chile. 

A set of three low resolution spectra, covering the wavelength range $3860-8070\,$\AA, were acquired on three different nights, one (2004 Mar 15), two (2004 May 1) and six (2004 Aug 27) months after the superoutburst, using grism \#\,6. Additional medium resolution spectra, covering the wavelength range $6280-8200\,$\AA, were obtained in March 2004 and August 2004 using grism \#10. These observations consisted of a series a consecutive spectra with exposure times varying between 300\,s and 400\,s.

The standard reduction of the data was performed using IRAF\footnote{IRAF is distributed by the National Optical Astronomy Observatories, which are operated by the Association of Universities for Research in Astronomy, Inc., under cooperative agreement with the National Science Foundation.} \citep{tody1986, tody1993}. The bias level has been subtracted and the data have been divided by a flat field, which was normalised by fitting Chebyshev functions of high order to remove the detector specific spectral response. All spectra have been optimally extracted following the method of \citet{horne86-1} and the low resolution spectra of each night have been averaged. Wavelength calibration yielded a final FWHM (full width at half--maximum) resolution of 12.8\,\AA\ for the low resolution data (grism \#\,6) and 5.5\,\AA\ for the medium resolution data (grism \#10). All further analysis has been done using the ESO--MIDAS toolkit.

For the low resolution data, the instrument response function was corrected using spectroscopic standards. Since the nights have been clear but not photometric, we performed differential photometry on the acquisition files to determine the relative flux values of our object with respect to four comparison stars. The spectra were scaled accordingly. Hence, while the absolute zero--point is more uncertain, the relative flux calibration of the individual spectra and the flux differences between the three average spectra have an accuracy of about 4\%.

High spectral resolution time--series spectroscopy was obtained with X--shooter at four epochs in 2012, 2013, and 2015. X--shooter is a medium--resolution {\'e}chelle spectrograph available at the Very Large Telescope (VLT) in Cerro Paranal (Chile, \citealt{xshooter}) since 2009. The instrument has been designed to cover in one exposure the wavelength range from $\simeq 3\,000\,$\AA\,up to $\simeq 25\,000\,$\AA. In order to do so, X--shooter is equipped with three arms: blue (UVB, $\lambda \simeq 3000-5595\,$\AA), visual (VIS, $\lambda \simeq 5595-10\,240\,$\AA) and near--infrared (NIR, $\lambda \simeq 10\,240-24\,800\,$\AA).  Spectra were obtained with slit widths of 1.0\arcsec\, in the UVB arm, 0.9\arcsec\, in the VIS arm, and 0.9\arcsec\, in the NIR arm and a $1\times2$ binning yielding spectral resolving powers of $R \sim 5000{-}9000$. 
The data processing was performed with the ESO Reflex X--shooter pipeline (version 2.4.0), which includes the standard reduction steps of bias and dark current removal, order identification and tracing, flat-fielding, dispersion solution, correction for instrument response and atmospheric extinction, and merging of all orders. 

QZ\,Lib is located at a galactic latitude of $36.4^\circ$ and at a distance of $d = 187 \pm 12\,$pc, as determined from the Gaia parallax ($\varpi = 5.3 \pm 0.3\,$mas, \citealt{Gaia2016,Gaia2018}). This implies an interstellar extinction of $E(B-V) = 0.09$, based on the three--dimensional map of interstellar dust reddening based on Pan--STARRS\,1 and 2MASS photometry \citep{panstar2017}. Such a small amount of interstellar extinction is negligible for the analysis presented here, and we hence show the spectra as observed, i.e. without reddening correction.

\begin{figure*}
\centering
\includegraphics[width=\textwidth]{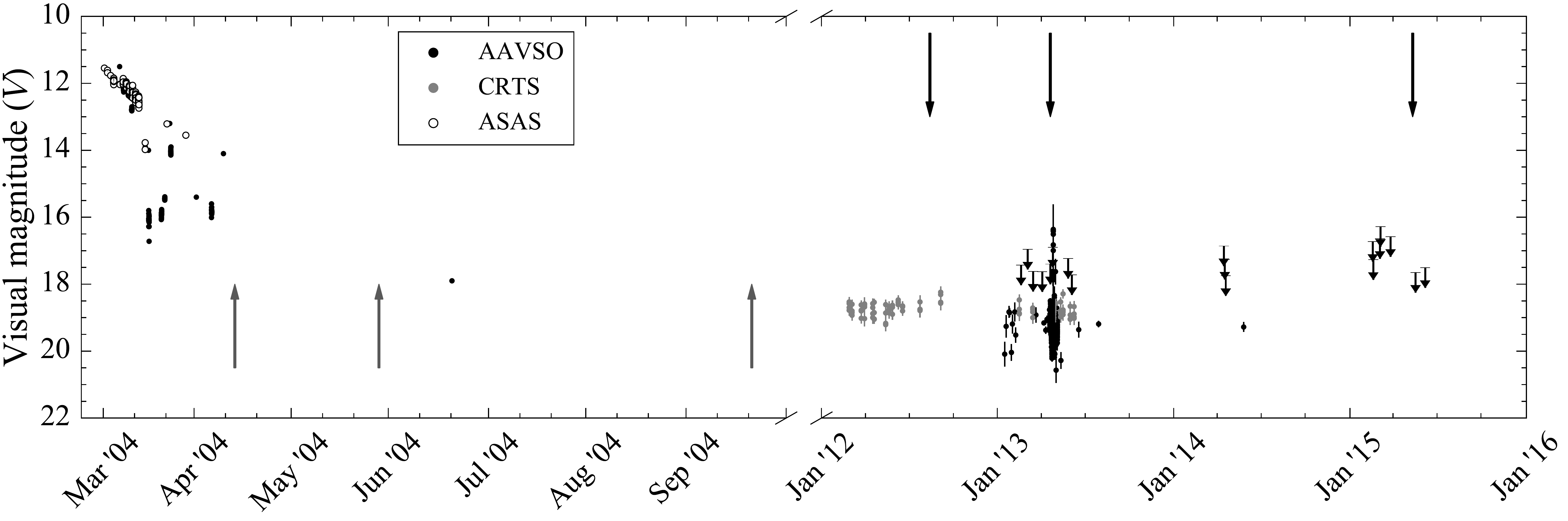}
\caption{\label{fig:aavso_lightcurve} Light curve of QZ\,Lib from its first detection (superoutburst in February 2004) up to 2016. The grey arrows highlight the epochs of the EFOSC2 observations one month (grism \#\,6 and \#\,10), two months (grisms \#\,6) and six months (grisms \#\,6 and \#\,10) after the superoutburst. The black arrows highlight the three epochs in which X--shooter spectroscopy has been acquired.}
\end{figure*}

\begin{figure*}
\centering
\includegraphics[width=\textwidth]{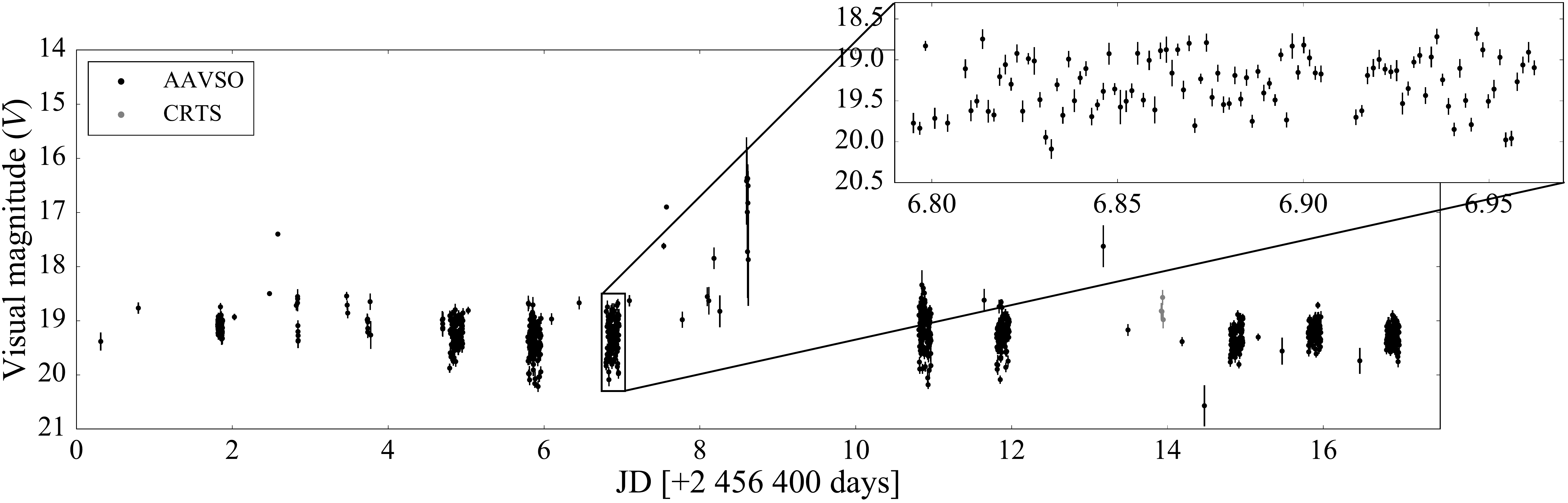}
\caption{\label{fig:aavso_lightcurve_zoom} Ligthcurve of QZ\,Lib in a period of particular intense monitoring (17 April 2013 -- 5 May 2013).
The inset shows a sample close--up of the night of 24--04--2013: the system variability is dominated by flickering and no periodic variation is detected.}
\end{figure*}

\subsection{Photometric observations}
Differential photometry at five minute cadence was performed using a $V$--band filter in front of a 512x512 CCD mounted on the 1.0\,m SMARTS (Small \& Moderate Aperture Research Telescope System) telescope at the Cerro Tololo Inter--American Observatory (CTIO), Chile, in February 2005. On the same month, 1.5\,h of photometry with higher time resolution ($20\,$s) were obtained with EFOSC2.
The details of the observations are given in Table~\ref{obstab}. The reduction was done with IRAF and included the usual steps of bias subtraction and division by skyflats. Aperture photometry for all stars on the CCD field was computed using the stand--alone version of DAOPHOT and DAOMASTER \citep{stetson92-1}. Differential light curves were established with respect to an average light curve of nine comparison stars (see Figure~\ref{chart}), which were present on all frames and were checked to be non--variable. 

\section{The light curve}
During the SMARTS and EFOSC2 observations, the brightness of QZ\,Lib is clearly varying, however, this variation is more of an irregular nature, with flickering and long--time variation dominating the light curves. We searched for periodic modulation using various Fourier techniques but did not find any persistent signal. 
Similar results can be derived from the photometric observations in the AAVSO (American Association of Variable Star Observers) and CRTS (Catalina Real-time Transient Survey, \citealt{Drake_et_al_2009}) databases (Figure~\ref{fig:aavso_lightcurve}): no orbital variation is detected even for those nights that are covered with ample data points (Figure~\ref{fig:aavso_lightcurve_zoom}). 

\section{Time resolved spectroscopy}\label{time_resolved}
\subsection{The orbital period}\label{subsec:orbital_period}
Time--resolved spectroscopic observations of QZ\,Lib were obtained with the EFOSC2 medium resolution grism \#\,10 on 2004 March 16-17 and 2004 August 28-30. Each set of observation covered about 1.5 orbits. The system had retuned to the quiescent state by the time of the August observations, as can be inferred from the additional low resolution EFOSC2 spectroscopic observations described in Section~\ref{sec:spectra}.

\begin{figure}
\centering
\includegraphics[angle=-90,width=0.48\textwidth]{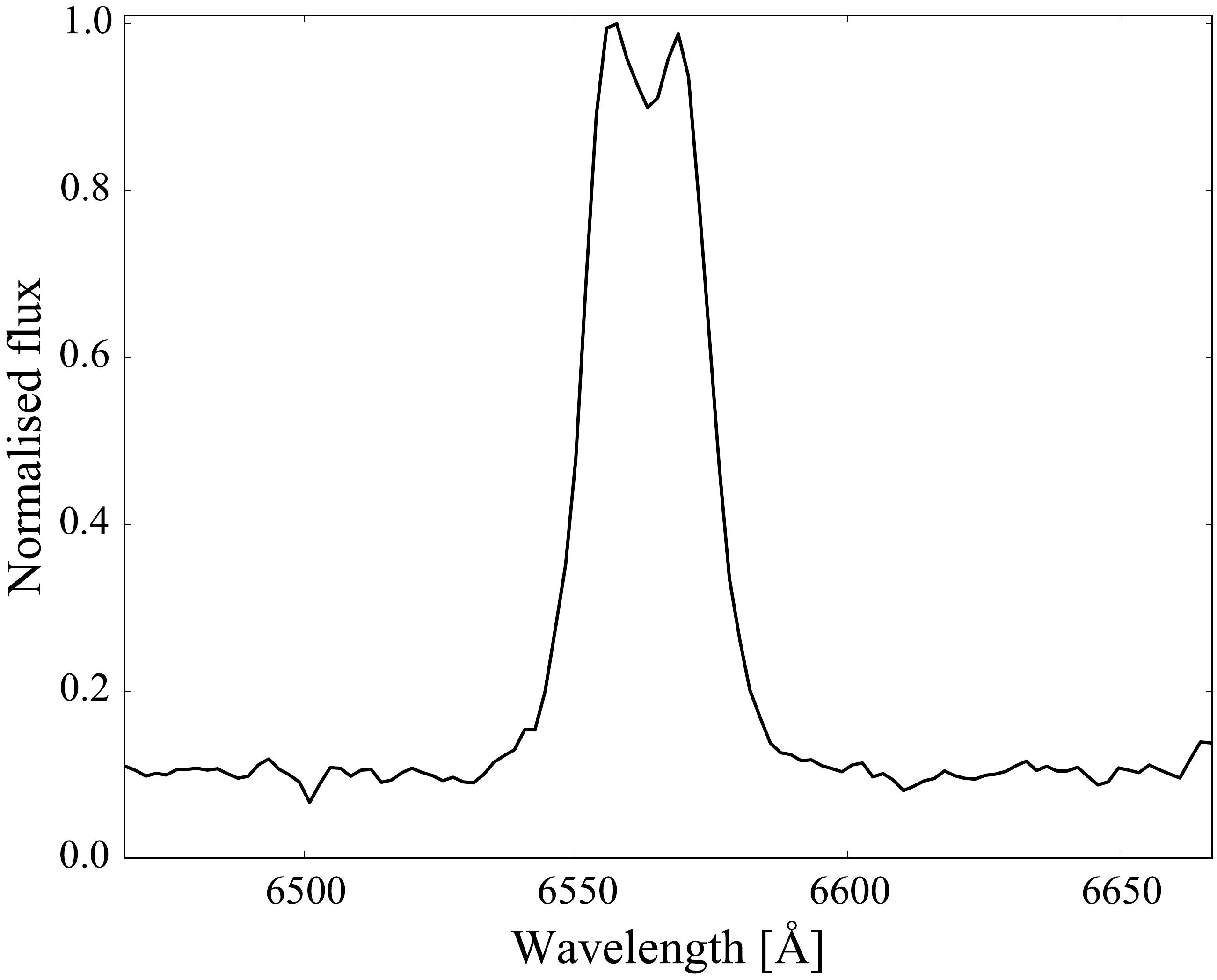}
\caption{EFOSC2 average spectrum of QZ\,Lib around the $H\alpha$ emission line, obtained with the grism \#\,10 six months after its 2004 superoutburst. As shown by the additional spectroscopic observations presented in Section~\ref{sec:spectra}, the system had returned to the quiescent level at the time of these observations.}\label{fig:efosc_ha}
\end{figure}

The average spectrum obtained during the August EFOSC2 grism \#\,10 observations is shown in Figure~\ref{fig:efosc_ha}. The H$\alpha$ emission line shows a double-peaked morphology, arising from the Keplerian velocity distribution of the gas in the disc \citep{marsh+horne88-1}. Since the disc is centred on the white dwarf, the Doppler shift of the H$\alpha$ line traces the motion around the centre of mass of the system and provides a measurement of the orbital period.
We therefore measured the radial velocity of the H$\alpha$ emission line by fitting a single broad Gaussian to it. 
A more complex fitting procedure, e.g. using multiple Gaussians \citep{shafter83-1}, may provide a better fit to the double-peaked structure, but will not result in a more accurate determination of the orbital period.

\begin{figure}
\centering
\includegraphics[angle=-90,trim={1.8cm 1.8cm 2cm 0.5cm},clip,width=0.48\textwidth]{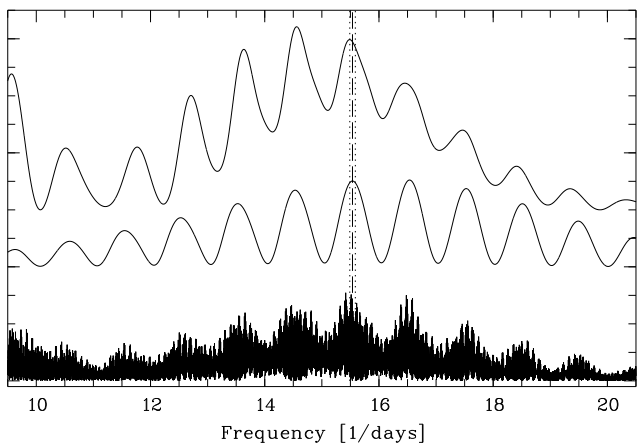}
\caption{Scargle periodograms for the March (top) and August (middle) EFOSC2 datasets. The power is shown on the $y$-axis, where an arbitrary scale is used for illustration purposes. The EFOSC2 datasets shows one strong peak in common, which we assumed to belong to the orbital period (vertical solid line and dashed lines for the related uncertainties). The same peak is also identified in the periodogram computed using all the EFOSC2 and the X--shooter observations (bottom). These observations are more than ten years apart and therefore the inclusion of the X--shooter data does not allow to improve the precision of the orbital period.}\label{fig:scargle}
\end{figure}

A clear periodicity of the H$\alpha$ emission line was found using Scargle and AOV Fourier techniques. The periodograms of both data sets show several alias peaks but only one frequency yields a strong signal in both, which we concluded to belong to the orbital period (see Figure~\ref{fig:scargle}). Since the August dataset has higher SNR and is slightly longer than the March dataset, we used it to determine the orbital period, which results in $P_\mathrm{orb}= 0.06436(20)$\,d or $1.545 \pm 0.005\,$h, in good agreement with the measurement by \citet{Thorstensen+2017}, $P_\mathrm{orb}= 0.06413(8)$\,d. Unfortunately, the March and August sets are too far apart in time to allow the combination of the data to increase the accuracy. In addition, some time--series spectroscopy at higher spectral resolution was taken with X--shooter during three epochs in 2012, 2013 and 2015 which we used to verify the orbital period. However, the longest X--shooter dataset from April 2013 covers less than two orbits and therefore does not allow to improve the precision of the orbital period. Moreover, the X--shooter and EFOSC2 data sets are too far apart in time to be combined and improve the uncertainty of the period determination.

\begin{figure}
\centering
\includegraphics[width=0.48\textwidth]{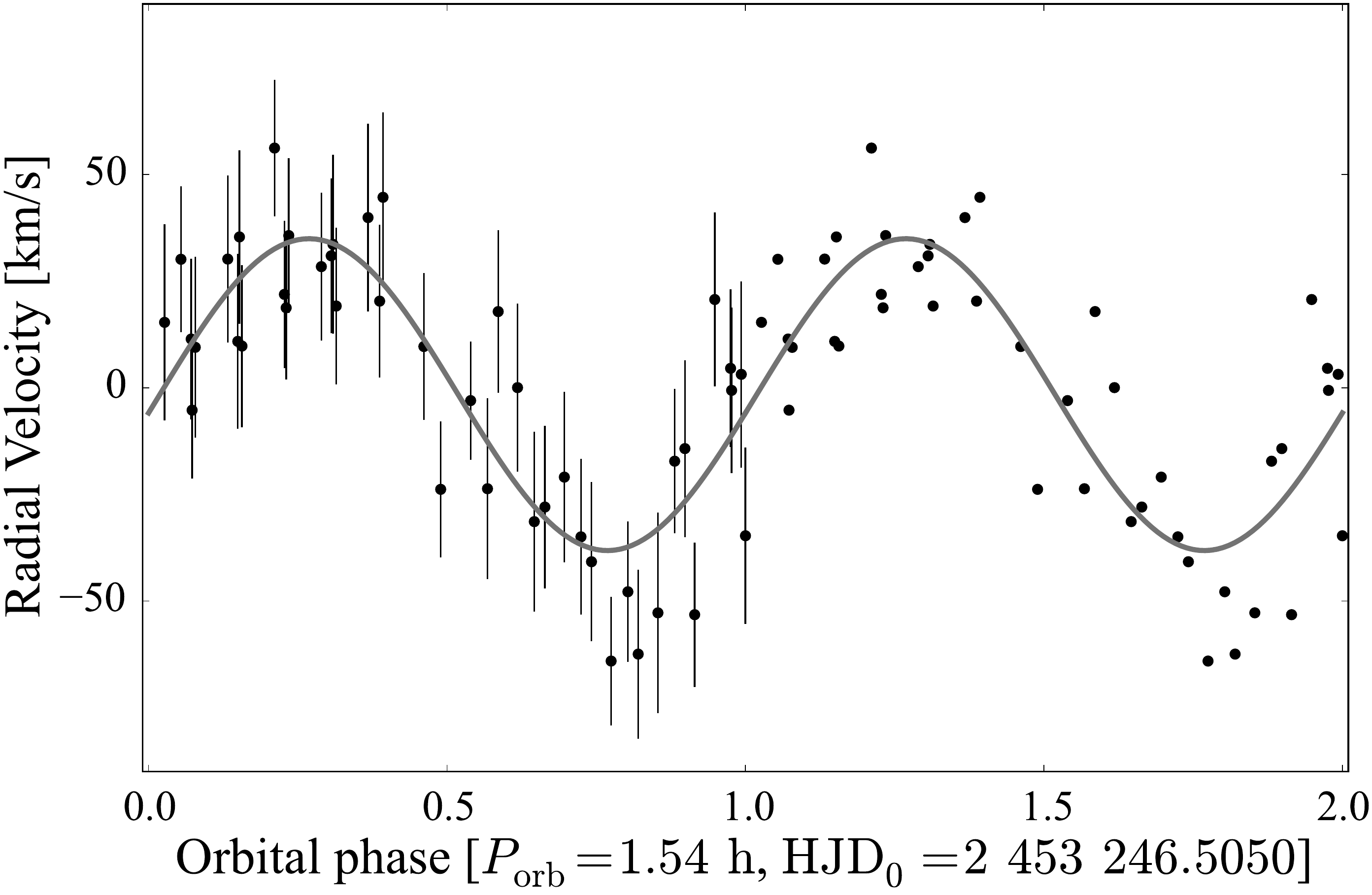}
\caption{The EFOSC2 August data are plotted in phase using the orbital period $P_\mathrm{orb}= 0.06436(20)$\,d; the zero-phase corresponds to $\mathrm{HJD} = 2\,453\,246.5050$. Two phases are plotted for clarity, the first one has the uncertainties of the individual measurements overplotted. The grey line gives the best sinusoidal fit to the data.}\label{lib_rvs}
\end{figure}

In a circular orbit, the radial velocity, $V$, of the $H\alpha$ emission line varies in a sinusoidal way during one orbital cycle:
\begin{equation}
\label{rad_vel}
V = \gamma + K_\mathrm{E} \sin (2 \pi \phi)
\end{equation}
where $\gamma$ is the systemic velocity of the system, $K_\mathrm{E}$ is the radial velocity amplitude of the emission lines and $\phi$ is the orbital phase of the spectrum.
Given the higher quality of the August EFOSC2 dataset, we used these time--resolved observations to compute $\gamma$ and $K_\mathrm{E}$. From a sinusoidal least squares fit to the data, we found $\gamma = -1.6 \pm 1.5$\,km\,s$^{-1}$ and $K_\mathrm{E} = 37 \pm 2$\,km\,s$^{-1}$. The uncertainties on these quantities were determined by computing Monte--Carlo simulations of the fit. 
The best fit returns the following ephemeris for the red to blue crossing of the emission lines:
\begin{equation}
\label{phi}
\mathrm{HJD} = 2\,453\,246.5050(6) + 0.06436(20)~E
\end{equation}
We then used this ephemeris to calculate the orbital phase for all data points of the August 2004 data set (Figure\,\ref{lib_rvs}).

\subsection{Doppler Tomography}
\begin{figure}
\centering
\includegraphics[width=0.48\textwidth]{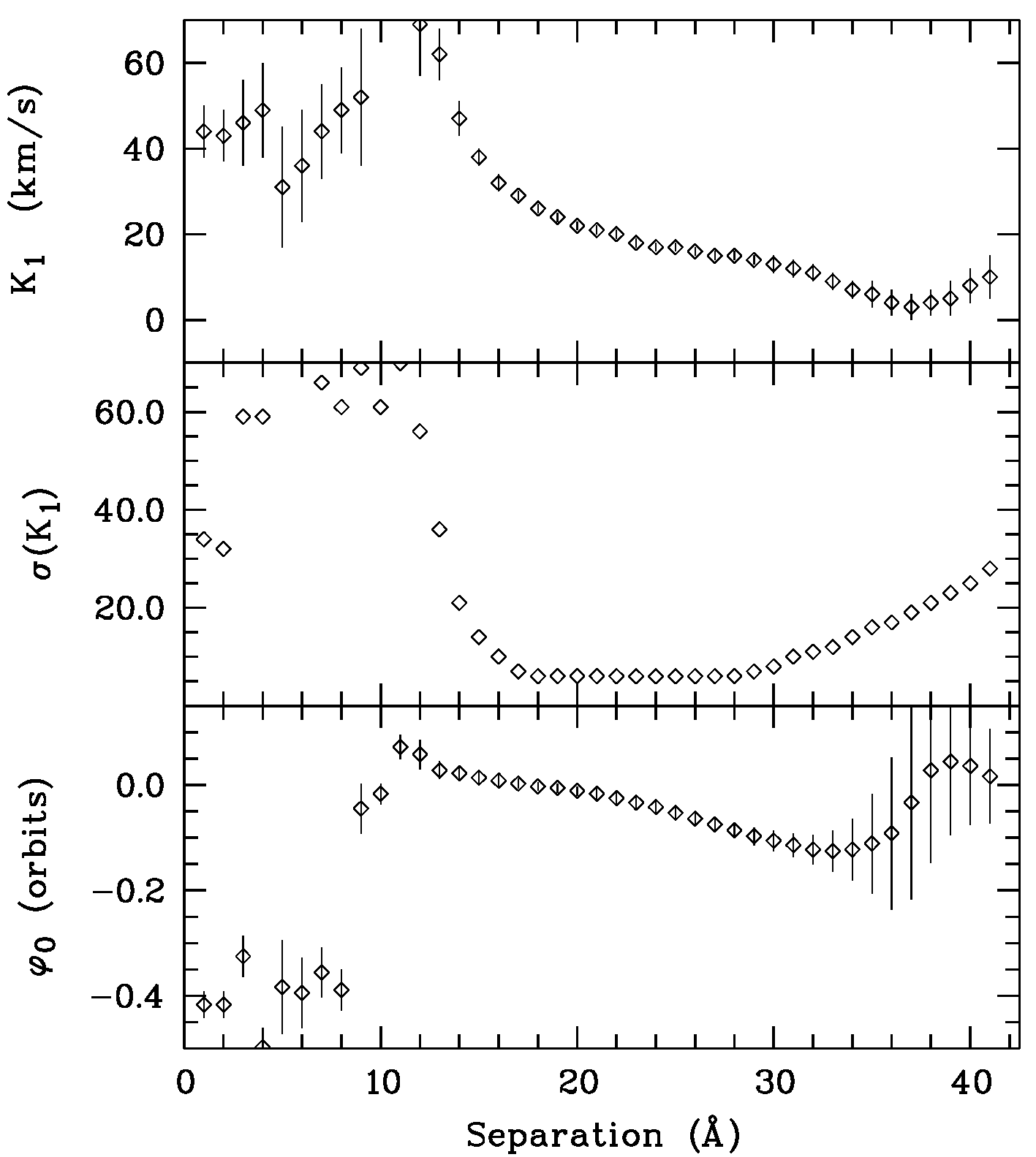}
\caption{Diagnostic diagram for the H$\alpha$ emission line, showing the parameters of the radial velocity fit as function of the Gaussian separation. For separations between $17\,$\AA\,and $28\,$\AA, where $\sigma(K_1)$ is minimal, the outer wings of the lines are sampled.}\label{diag}
\end{figure}

Doppler tomography makes use of the orbital variation of the emission line profiles. Each point of the emission line can be attributed to a radial velocity, which can be transformed to $(v_x,v_y)$ for a given phase. The Doppler maps $I(v_x, v_y)$ display the flux emitted by gas moving with the velocity $(v_x , v_y)$ and thus show the emission distribution in velocity coordinates with the centre of mass at (0,0). We used the code of \citet{spru98} with a MIDAS interface replacing the original IDL routines \citep{tappertetal03-1} and the largest contiguous data set with the highest spectral resolution -- the X--shooter data from April 2013 -- for these computations. 

\begin{figure}
\centering
\includegraphics[width=0.48\textwidth]{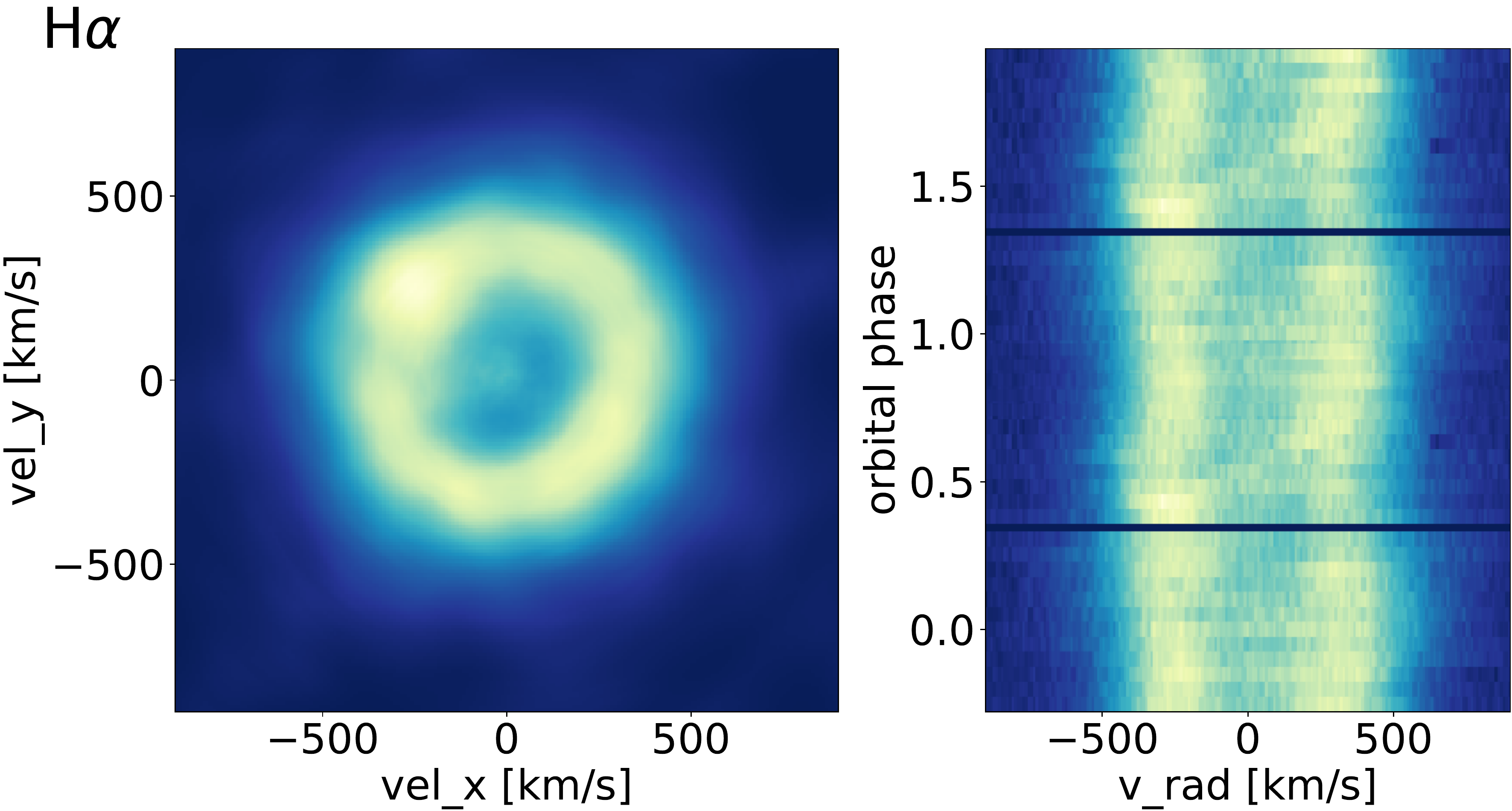}\\
\includegraphics[width=0.48\textwidth]{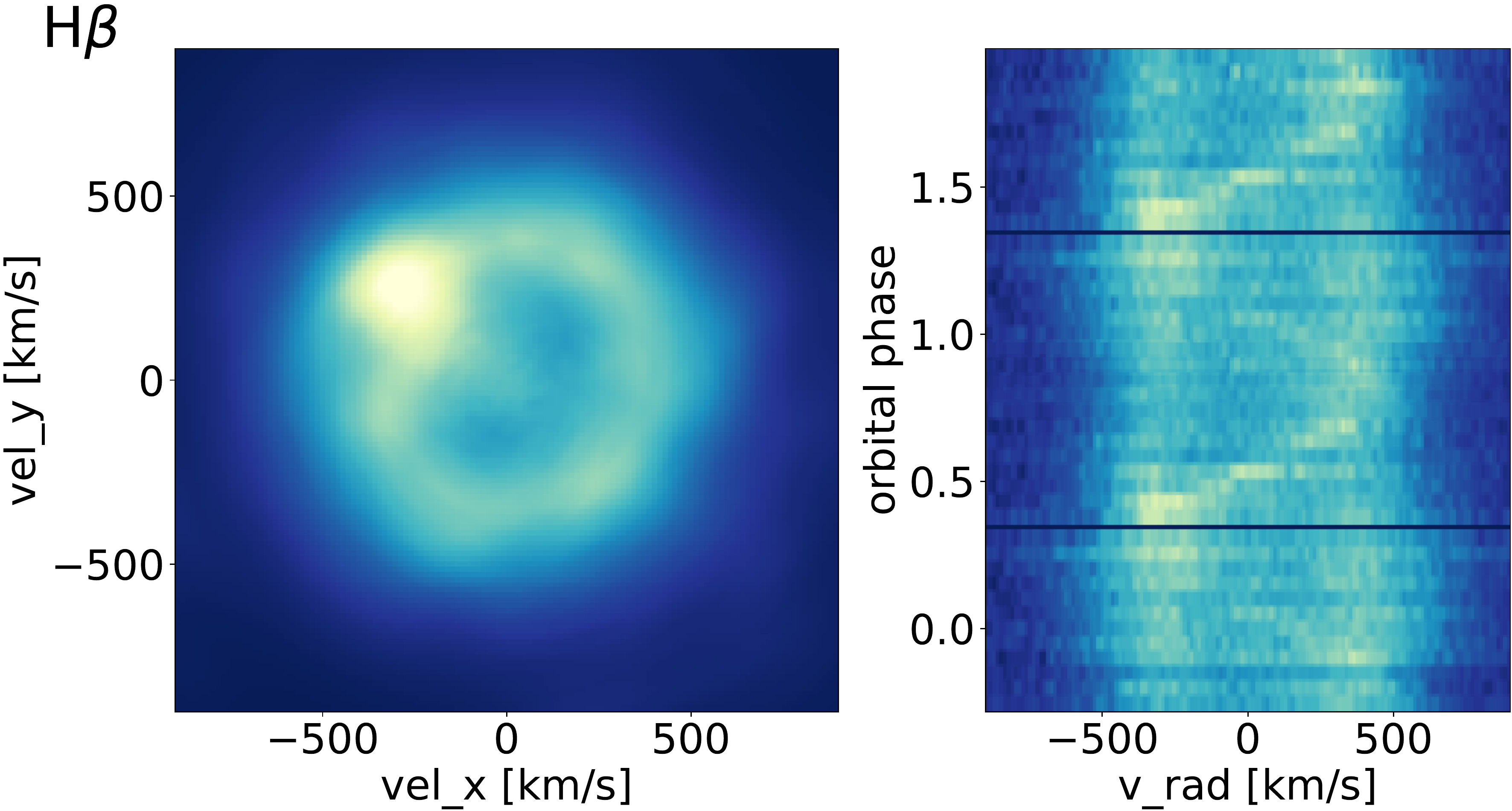}
\caption{The Doppler maps show the emission line distribution in velocity coordinates for H$\alpha$ (upper plot) and H$\beta$ (lower plot). In the panel on the right, the trailed spectra of the respective emission line are plotted in velocity coordinates. Two phase cycles are plotted for clarity.}\label{dopHaHb}
\end{figure}

The spectral resolution of the X--shooter data ($R \simeq 7000$) allows to accurately trace the motion of different parts of the disc which can be used to calculate the diagnostic diagram of the system \citep{shafter83-1} and better constrain the Doppler maps.
This method allows to measure the orbital velocity of the white dwarf ($K_\mathrm{WD}$) by probing the wing of the emission lines. This because the wings arise from the inner region of the accretion disc and therefore trace more accurately the motion of the white dwarf than the core of the emission line, which instead originates in the outer disc.
In the case of the line emission originating exclusively in a perfectly circular disc, the full emission line profile would reflect the motion of the white dwarf, i.e. $K_\mathrm{E} = K_\mathrm{WD}$. Instead, the symmetry of the outer disc is usually perturbed by the impact of the accretion stream, and additional emission components from this region and/or the potentially irradiated secondary star often contribute to the centre of the emission line, so that different parts of the emission line may correspond to different radial velocity amplitudes and phases \citep[e.g.][]{horne+marsh86-1}. In consequence, for measurements of the full line profile, usually $K_\mathrm{E} \neq K_\mathrm{WD}$.
 
We therefore measured $K_\mathrm{WD}$ in an iterative way, adopting the orbital period as determined from the analysis of the EFOSC2 data (see Section~\ref{subsec:orbital_period}) and using two Gaussians of 4\,\AA\ of FWHM to fit the $H\alpha$ emission lines in the phase resolved X--shooter observations from April 2013. The separation of the two Gaussians was varied between 1\,\AA\,and 42\,\AA\ to gradually exclude the core of the line.
For each separation, a radial velocity curve was produced and fitted with a sinusoidal function as defined in Equation~\ref{rad_vel}. The resulting velocity amplitudes $K_1$ as a function of the separations of the two Gaussians are shown in Figure~\ref{diag}.
Once the separation of the Gaussians is sufficient to exclude the perturbed core of the emission line, the velocity amplitude and the orbital phase remain roughly constant, and the most robust measurements of these two parameters is taken from the range of separations where $\sigma(K_1)$ is minimal.
In this case, this corresponds to separations between 17\,\AA\,and 28\,\AA, consistent with those values where the uncertainty $\sigma(K_1)$ has a minimum. To determine the zero-phase as input for the Doppler tomography, as well as the radial velocity amplitude of the white dwarf, we calculated the average values for this range of separations. For the white dwarf, we thus derived $K_\mathrm{WD} = 20 \pm 4 \,\rm km\,s^{-1}$.

The final Doppler maps for H$\alpha$ and H$\beta$ are shown in Figure~\ref{dopHaHb}, illustrating the distribution of the emission sources in velocity coordinates. To correct for the variation of the equivalent width, the input spectra have been normalised by the respective emission line flux. The intensity values in the Doppler map are hence to be interpreted as relative flux values only. The orientation of the map is chosen in such a way that the phase angle $\phi_r$ with respect to the zero point of radial velocities as determined from the diagnostic diagram is zero towards the top and increases clockwise. Apart from the disc itself, which is clearly visible, a significant additional source of emission can be seen in the upper left quadrant, located at a phase angle $\phi_r \approx 0.75$, which can be interpreted as emission from the bright spot, the point where the accretion stream hits the disc. The emission from the bright spot is also visible as a thin S--curve with a velocity amplitude consistent with the outer disc velocity in the trailed spectra shown in the right--hand side panels.

The presence of this isolated emission source explains the variation of $K_1$ within the diagnostic diagram as well as the difference between $K_E = 37\, \rm kms^{-1}$ as determined from fitting a broad Gaussian and $K_\mathrm{WD} = 20\, \rm kms^{-1}$ as determined from the diagnostic diagram. The fit of a single broad Gaussian to the complicated structure of the emission line is influenced by the isolated emission source and does not represent the motion of the white dwarf.

\subsection{The mass ratio}\label{subsec:mass_ratio}
From the orbital period $P_\mathrm{orb}= 0.06436(20)$\,d and the reported superhump period $P_\mathrm{SH} = 0.064602(24)$\,d \citep{kato15-1}, we derived a period excess of $\epsilon = (P_\mathrm{SH} - P_\mathrm{orb})/P_\mathrm{orb} = 0.0038$. Using the empirical formula $\epsilon = 0.18 q + 0.29 q^2$ \citep{pattersonetal05-3}, we find the mass ratio $q = 0.020 \pm 0.017$. This value is consistent with the previously determined one by \citeauthor{pattersonetal05-2} (\citeyear{pattersonetal05-2}, $q = 0.035 \pm 0.020$). We consider our measurement a more reliable result since their orbital period was tentatively derived from photometric variations (Patterson, private communication) which we consider problematic due to strong non--periodic contributions like flickering and long-term variations.

Since $P_\mathrm{SH}$ varies with time, the $\epsilon - q$ relationship can be applied at different stages (i.e. using different superhump periods) of the superoutburst. \cite{katoetal09-1} identified three superhump stages: (A) an early phase where $P_\mathrm{SH}$ has the highest value; (B) an intermediate phase with a stabilised $P_\mathrm{SH}$ and (C) a final phase with a shorter $P_\mathrm{SH}$. In the traditional way of determining $q$ from the period excess \citep{pattersonetal05-3,katoetal09-1,Patterson2011}, $P_\mathrm{SH}$ determined during stage B is commonly used. However, \citet{kato2013} argue that this $P_\mathrm{SH}$ systematically underestimates $q$ for low mass--ratio systems ($q \lesssim 0.09$) and they present an alternative $\epsilon - q$ relationship calibration from stage A superhumps. The mass--ratio we derived for QZ\,Lib lies in the range affected by this systematic effect and therefore we re--computed the value of $q$ using equation 5 from \citet{kato2013}, using the stage A superhump period of QZ\,Lib from \citeauthor{katoetal2015} (\citeyear{katoetal2015}, $P_\mathrm{SH} = 0.06557(14)$\,d), finding $q = 0.048 \pm 0.010$.

The mass ratios derived with the two $\epsilon - q$ calibrations, respectively $q = 0.020 \pm 0.017$ (obtained using the method by \citealt{pattersonetal05-3}) and $q = 0.048 \pm 0.010$ (obtained using the method by \citealt{kato2013}) agree within the uncertainties (on a $\simeq 1.4\,\sigma$ level). Assuming one instead of the other does not affect the conclusions drawn in the following Sections, in which we assume the weighted average of the two: $q = 0.040 \pm 0.009$.

\section{The spectra}\label{sec:spectra}
\begin{figure}
\centering
\includegraphics[width=0.48\textwidth]{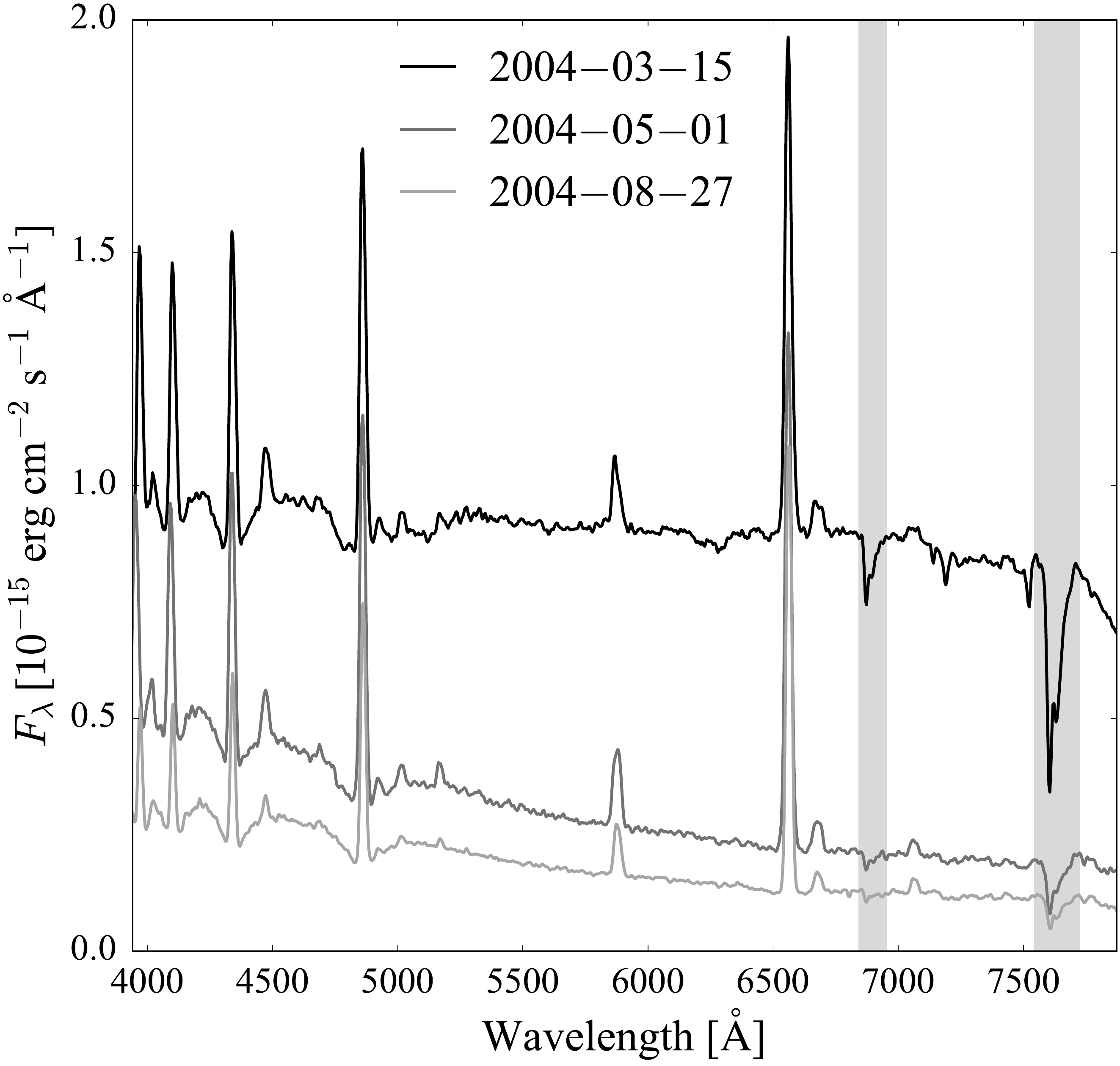}
\caption{EFOSC2 spectrophotometric calibrated spectra of QZ\,Lib obtained one (black), two (dark grey) and six (light grey) months after its 2004 superoutburst. The regions contaminated by telluric absorption bands are highlighted in grey.}\label{spectra_lr}
\end{figure}

\begin{table}
\centering
\caption{\label{table:free_params} Range of variations for the free parameters in the slab model describing the emission from accretion disc.} 
\begin{tabular}{l l}
\toprule
 \noalign{\smallskip}
 Parameter & Value \\
\midrule
 \noalign{\smallskip}
Effective temperature (K) & $5700 - 8000$ \\
Pressure (dyn\,cm$^{-2}$) & $0 - 1000$ \\
Radial velocity (km\,{s}$^{-1}$) & $0 - 2000$\\
Inclination (\degree) & $0 - 60$\\
Geometrical height (cm) & $0 - 10^{10}$\\
 \noalign{\smallskip}
\bottomrule
\end{tabular}
\begin{tablenotes}
\item \textbf{Notes.} The upper limit on the inclination is set by the fact that no eclipse is detected in the light curve of QZ\,Lib. \\
\end{tablenotes}
\end{table}

\subsection{The model}\label{sec:model}
In the optical spectrum of a CV, three main light sources contributes to the emission of the system: the white dwarf, the secondary star and the accretion disc. The white dwarf, which is heated by the energy released by the compression of the accreted material \citep{sion95-1,townsley+bildsten03-1}, is usually hot with an effective temperature $T_\mathrm{eff} \gtrsim 10\,000\,$K \citep{sion99-1}, and thus dominates in the blue part of the spectrum. The donor is typically a cold low--mass star which starts contributing in the near--infrared ($\lambda \gtrsim 7000\,$\AA). Finally, strong emission lines, whose shape and width are determined by the Keplerian velocity distribution of the accreting material, reveal the presence of the disc.

From a spectral fit to a CV spectrum with synthetic models accounting for the contribution from these three sources, it is possible to determine the effective temperatures of the white dwarf and the secondary star. We used \textsc{tlusty} and \textsc{synspec} \citep{tlusty,tlusty1} to compute a grid of white dwarf atmosphere models covering the range $T_\mathrm{eff} = 9000 -  40\,000\,$K in steps of $100\,$K and assuming a metallicity of $Z = 0.01\,Z_\odot$, (as determined from the analysis of ultraviolet \textit{Hubble Space Telescope} data of QZ\,Lib by \citealt{Pala}).
The white dwarf effective temperature correlates with its surface gravity: strong gravitational fields translate into pressure broadening of the lines; this effect can be balanced by higher temperatures that increase the fraction of ionised hydrogen, resulting in narrower absorption lines. Moreover, in CVs, the Balmer line cores, commonly used to simultaneously constrain the effective temperature and the surface gravity of isolated white dwarfs (see for example \citealt{Gianninas2011}), are often contaminated by strong disc emission lines. For these reasons, it is not possible to constrain both $T_\mathrm{eff}$ and $\log g$ of the white dwarf in QZ\,Lib from the analysis of optical data and we therefore generated our grid of models assuming $\log g = 8.35$, corresponding to the average mass of CV white dwarfs ($M_\mathrm{WD} = 0.83 \pm 0.23\,\mathrm{M}_\odot$, \citealt{Zorotovic}). Assuming the mass--radius relationship from \citet{m-r_relationship}, this value of $\log g$ correspond to a white dwarf radius of $R_\mathrm{WD} = 0.01\,\mathrm{R}_\odot$ (the assumption of the zero-temperature mass--radius relationship is sufficient given the low temperature of the white dwarf).

As discussed by \citet{Pala}, the correlation between the white dwarf $\log g$ and $T_\mathrm{eff}$ introduces a systematic uncertainty in the effective temperature measurement (see figure~8 from \citealt{Pala}) which is of the order of $\simeq 200\,$K for $T_\mathrm{eff} \simeq 11\,000\,$K and increases up to $\simeq 1000\,$K for $T_\mathrm{eff} \simeq 17\,000\,$K. In order to account for its effect, in the following, we assumed as uncertainty on our effective temperature determinations the largest between (i) the statistical uncertainty returned by our fitting procedure and (ii) the corresponding systematic uncertainty related to the unknown mass of the white dwarf (as determined from equation~2 from \citealt{Pala}). In this way, our analysis accounts also for the uncertainty of the mass distribution of CV white dwarfs.

\begin{table}
\centering
\caption{\label{table:cooling} Cooling sequence for the white dwarf in QZ\,Lib after its 2004 super--outburst.} 
\begin{tabular}{l l}
\toprule
 \noalign{\smallskip}
 Date & $T_\mathrm{eff}$ (K) \\
\midrule
 \noalign{\smallskip}
2004--03--15 & $17\,000 \pm 2000$ \\
2004--05--01 & $14\,000 \pm 600 $ \\ 
2004--08--27 & $11\,700 \pm 250 $ \\
 \noalign{\smallskip}
\bottomrule
\end{tabular}
\end{table}

\begin{figure}
\center
\includegraphics[width=0.48\textwidth]{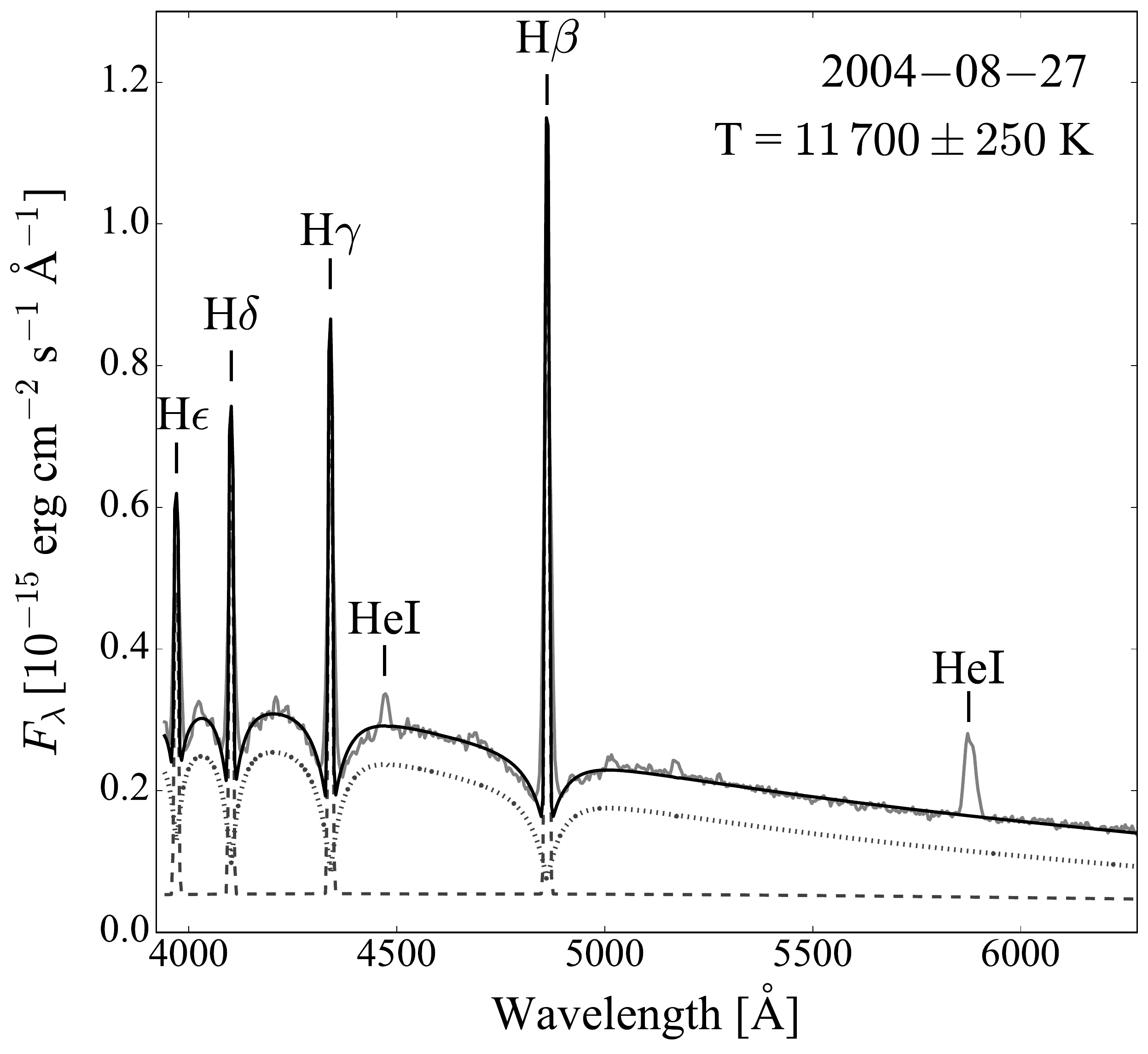}
\caption{EFOSC2 spectrum of QZ\,Lib (light grey) taken on 2004--08--27. The \#6 grism is affected by a second--order contamination which reduces the useful wavelength range to $6200\,$\AA. At these wavelengths the possible contribution of the donor is negligible and thus we fitted the data including the contribution of a white dwarf (dotted line) and an isothermal and isobaric pure hydrogen slab (dashed line) to approximate the disc emission. The sum is given by the black solid line.}\label{lib_wd_2}
\end{figure}

To approximate the disc emission, we used an isothermal and isobaric pure--hydrogen slab model, as described in \citet{Boris_disc,Boris_disc1}. The free parameters of this model and their allowed ranges are reported in Table~\ref{table:free_params}.
 
 \begin{figure*}
\center
\includegraphics[angle=-90,width=0.8\textwidth]{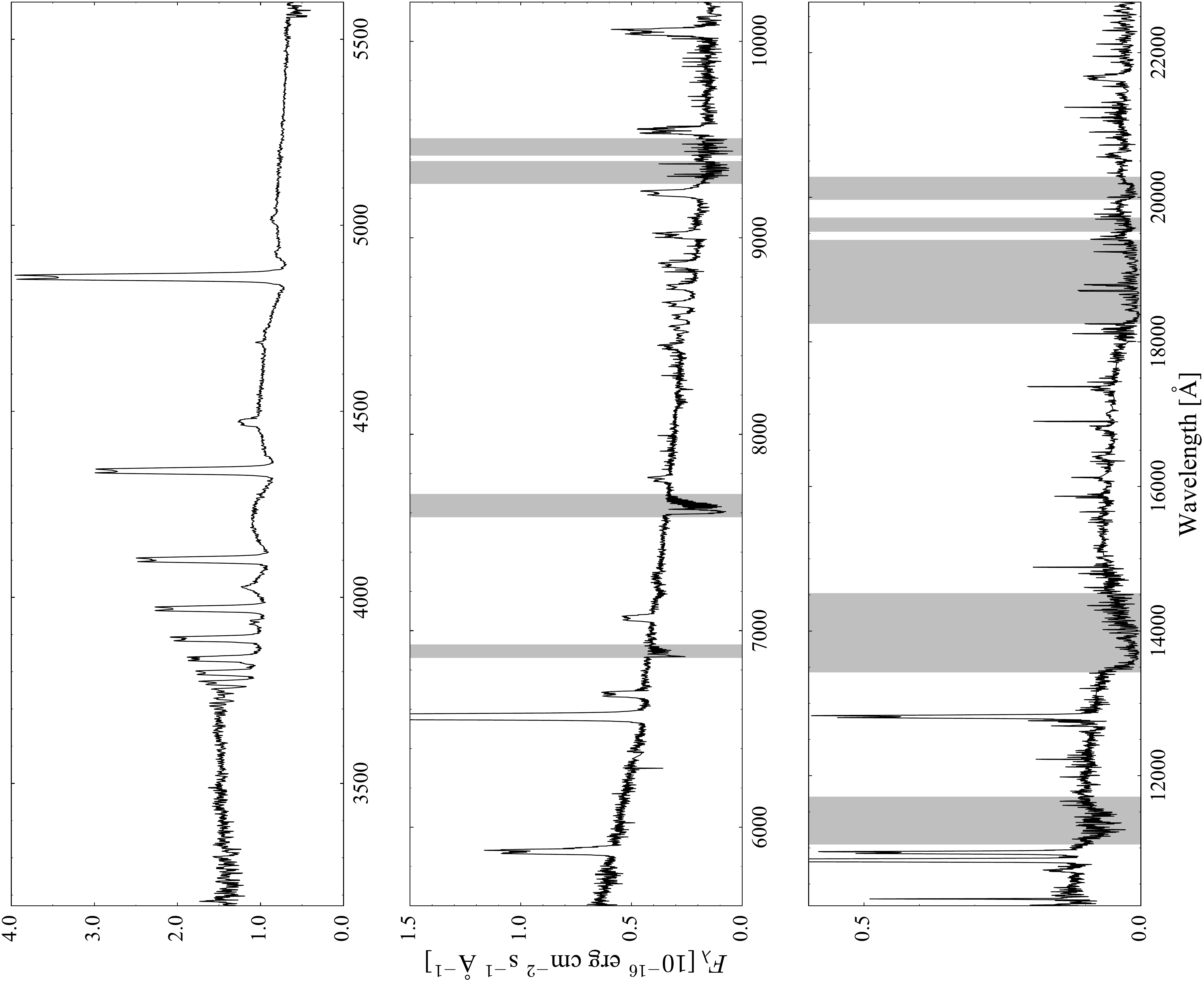}
\caption{Averaged X--shooter spectrum of QZ\,Lib (UVB, top panel; VIS middle panel; NIR, bottom panel). The white dwarf signature is recognisable in the broad Balmer lines in the UVB arm. Strong emission lines arise from the accretion disc. No features from the secondary are detected in the near--infrared, which is contaminated by broad telluric absorption bands (highlighted in grey) and residuals from the sky line removal.}\label{xshooter_ave}
\end{figure*}

Finally, the last contribution that needs to be taken into account is the emission from the donor star. 
From the knowledge of the system mass ratio and from our previous assumption of $M_\mathrm{WD} = 0.8\,\mathrm{M}_\odot$, we determined the mass of the secondary star, $M_\mathrm{sec} = 0.032 \pm 0.012\,\mathrm{M}_\odot$. 
The orbital period of the system and Kepler's third law return a measurement of the orbital separation,  $a = 0.63 \pm 0.05\,\mathrm{R}_\odot$. Since the secondary is filling its Roche--lobe, its radius is defined by the relationship \citep{sec_radius}: 
\begin{equation}
R_\mathrm{sec} = \frac{a\,0.49\,q^{2/3}}{0.6\,q^{2/3} + \ln (1\,+\,q^{1/3})}
\end{equation}
and results in $R_\mathrm{sec} =  0.10 \pm 0.01\,\mathrm{R}_\odot$. The combination $M_\mathrm{sec}$ and $R_\mathrm{sec}$ returns the surface gravity of the donor, $\log(g[\mathrm{cm}\,\mathrm{s}^{-2}]) = 4.9 \pm 0.2$. The typical effective temperature of donor stars in short--period CVs are $500\,\mathrm{K} \lesssim T_\mathrm{eff} \lesssim 3000\,\mathrm{K}$ \citep{Knigge2006}. The metallicity of the white dwarf represents a lower limit on the metallicity of the accreted material, which is stripped from the secondary photosphere. We therefore assumed that the secondary has the same metallicity as the white dwarf, $Z = 0.01\,Z_\odot$. However, at present, none of the available model grids for late--type stars for $Z = 0.01\,Z_\odot$ cover the required range in temperature. We therefore retrieved a grid of BT--Dusty \citep{BT-dusty} models for late--type stars from the Theoretical Spectra Web Server\footnote{\href{Theoretical Spectra Web Server}{http://svo2.cab.inta-csic.es/theory/newov/index.php?model=bt-settl}}, extending from $T_\mathrm{eff} = 1000\,$K (T6 spectral type) up to $T_\mathrm{eff} = 3600\,$K (M2 spectral type), for $\log g = 5$ and $Z = 0.3\,Z_\odot$, which is the widest grid of available models with the lowest metallicity value.

Fixing the distance to the system to $d = 187\,$pc, as derived from the Gaia parallax, we then used a $\chi^2$ minimisation routine to fit our grid of models to the optical spectra. In the fitting procedure, we did not include the wavelength ranges of the telluric absorption features arising in the Earth's atmosphere and, to better constrain the contribution of the disc to the overall emission, we allowed as free parameters the areas of the emission lines H$\beta$, H$\gamma$, H$\delta$ and H$\epsilon$. In the case of the X--shooter spectra we included also the areas of the emission lines H$\alpha$, Pa$\delta$ and Pa$\zeta$. 

We present the results of the fit to the different datasets in the following Sections.

\begin{table*}
\caption{The evolution of FWHM (in \AA), equivalent width $W$ (in \AA), and line flux $F$ (in $\rm10^{-15} erg\, cm^{-2} s^{-1}$) for the main Balmer and \ion{He}{i} emission lines in the spectra of QZ\,Lib. Note that the uncertainty of the line flux accounts for the uncertainty of the relative flux in the line but does not include the errors introduced in our spectrophotometric flux calibration procedure (see Section~\ref{subsec:spec_obs}).}\label{tab:lines}
\begin{tabular}{l r r r c r r r} 
\toprule
          & \multicolumn{3}{c}{2004-03-15} & ~ & \multicolumn{3}{c}{2004-05-01}\\
Transition & FWHM  & \multicolumn{1}{c}{$-W$} & \multicolumn{1}{c}{$F$} 
& & FWHM & \multicolumn{1}{c}{$-W$} & \multicolumn{1}{c}{$F$} \\
\midrule
 H$_\alpha$ & $28.6\pm0.2$ & $38.7\pm0.8$ & $35.3\pm0.3$ & & $26.8\pm0.2$ & $157\pm1.5$ & $34.0\pm0.3$\\
 H$_\beta$  & $26.2\pm 0.3$ & $28.1\pm0.4$ & $24.9\pm0.3$ & & $24.3\pm0.3$  &  $73.5\pm0.8$ & $23.2\pm0.3$\\
 H$_\gamma$ & $25.1\pm0.3$& $21.7\pm0.3$ &  $19.0\pm0.3$ & & $23.6\pm0.4$ &  $40.8\pm0.9$ & $16.4\pm0.3$\\
 H$_\delta$ &$24.5\pm0.2$ & $16.7\pm0.3$ &  $15.4\pm0.2$ & & $23.6\pm0.4$ &  $28.1\pm0.9$ & $13.0\pm0.5$ \\
 \ion{He}{i}\,$\lambda$6678 & $46\pm 3$& $3.4\pm0.6$ & $3.0\pm0.2$ & &$39.0\pm0.5$ & $13.6\pm0.5$ & $2.9\pm0.2$ \\
  \ion{He}{i}\,$\lambda$5875 & $35\pm 2$& $6.2\pm0.8$ & $6.2\pm0.3$ & & $34.3\pm0.6$ & $23.3\pm0.8$ & $6.3\pm0.4$ \\
 \ion{He}{i}\,$\lambda$4471 &$34\pm 3$ & $5.1\pm0.4$ & $4.7\pm0.2$ & & $31.5\pm0.8$ &  $9.7\pm0.5$ & $4.3\pm0.3$ \\
\midrule
\midrule 
           & \multicolumn{3}{c}{2004-08-27} & ~ & \multicolumn{3}{c}{X--shooter average (2011-2015)} \\
Transition & FWHM  & \multicolumn{1}{c}{$-W$} & \multicolumn{1}{c}{$F$} 
& & FWHM & \multicolumn{1}{c}{$-W$} & \multicolumn{1}{c}{$F$} \\
\midrule
 H$_\alpha$ & $24.1\pm0.1$ & $214\pm2.5$ & $27.1\pm0.3$ & & $21.4\pm0.1$ & $288\pm7$ & $12.8\pm0.1$ \\
 H$_\beta$ & $20.6\pm0.2$ &  $69.1\pm0.9$ &  $13.6\pm0.2$& & $17.5\pm0.2$ & $82\pm3 $& $5.6\pm 0.1$ \\
 H$_\gamma$ & $18.8\pm0.3$ &  $33.5\pm0.7$ &  $8.1\pm0.2$& &  $16.4\pm0.5$ & $43\pm 2$& $3.5\pm0.1$ \\
 H$_\delta$ & $17.4\pm0.4$ &  $21.7\pm0.5$ &  $5.5\pm0.2$& & $15.6\pm0.4$ &$26\pm 2$ & $2.3\pm0.1$ \\
 \ion{He}{i}\,$\lambda$6678 & $33.6\pm0.2$ & $14\pm0.9$ &  $1.9\pm0.2$& & $26.5\pm0.4$ & $13\pm 1$ & $0.6\pm0.1$\\
 \ion{He}{i}\,$\lambda$5875 & $30.2\pm 0.2$ & $23\pm0.9$ & $4.1\pm0.3$& & $25.5\pm0.2$ &$23\pm2$ & $1.3\pm0.1$\\
 \ion{He}{i}\,$\lambda$4471 &$26\pm3.5$ &  $7\pm1.5$ & $1.7\pm0.2$& & $18.2\pm0.5$ & $4\pm 1$ & $0.5\pm0.1$\\
\bottomrule
\end{tabular}
\end{table*}

\begin{figure*}
\centering
\includegraphics[width=0.82\textwidth]{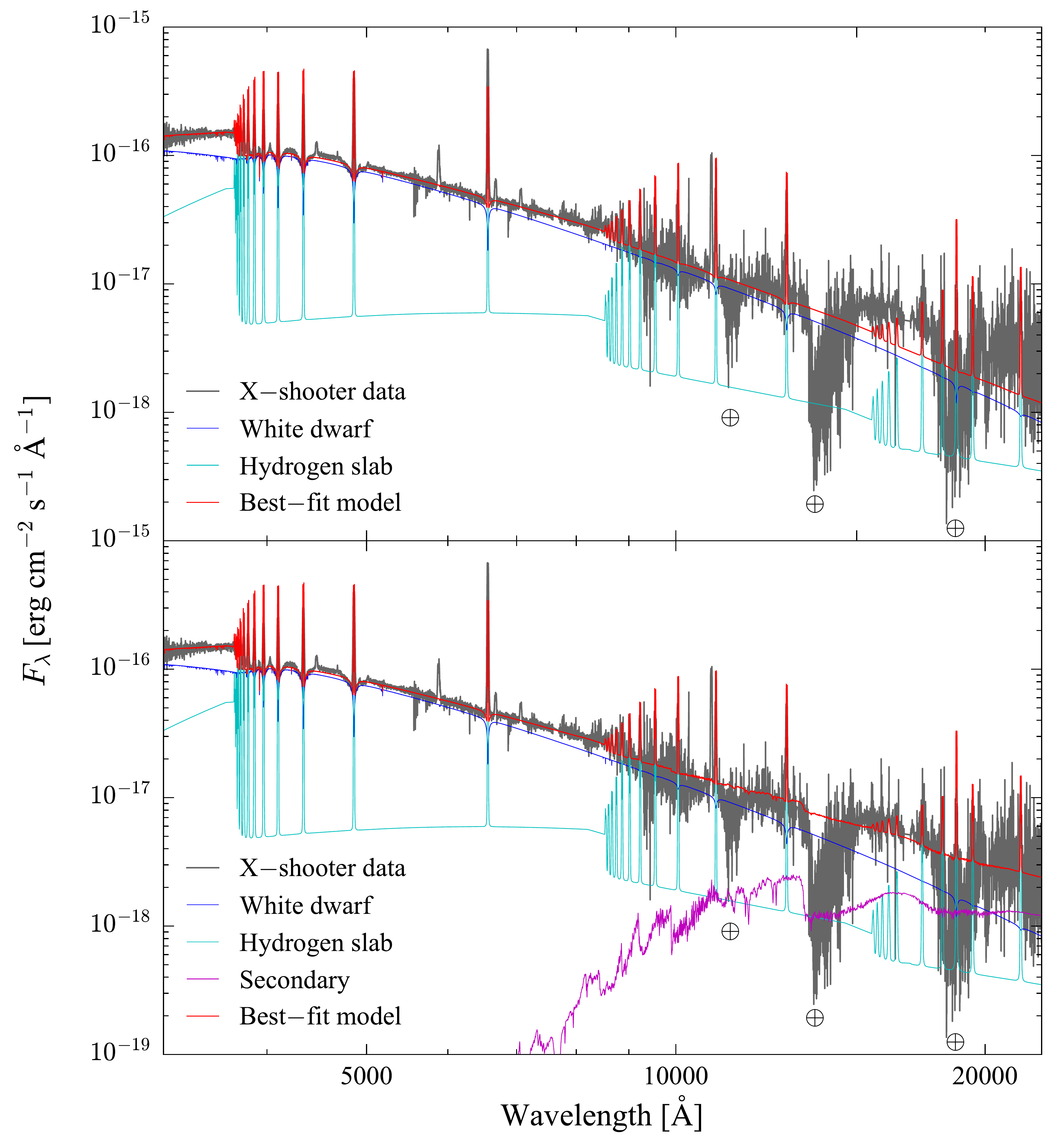}
\caption{X--shooter average spectrum of QZ\,Lib (grey) along with the best--fit model (red) which is composed of a white dwarf (blue) and an isothermal and isobaric hydrogen slab (cyan) and a brown dwarf companion (magenta, bottom). The Earth's symbols highlight the position of the strongest telluric absorption bands, which were not included in the fit. A colour version of this figure is available in the electronic version of the manuscript.}\label{xshooter_fit_sec} 
\end{figure*}

\subsection{The spectra during the cooling towards quiescence}
The spectrophotometric calibrated low resolution EFOSC2 spectra of three epochs are plotted in Figure~\ref{spectra_lr}. The white dwarf is dominating the continuum in all of them, as seen from the clear presence of its broad Balmer absorption lines. The spectrum taken on 2004--03--15 still has a noticeable contribution of the accretion disc which then decreases at the later epochs. 

Unfortunately, we noticed after the observations that the grism \#\,6 of EFOSC2 is affected by a second--order contamination starting around 6000\,\AA . Hence, no conclusions can be made for the cool donor star and only the blue part of the spectrum was used for the fit of the white dwarf and the accretion disc.

Following the methodology described in the previous section, we fitted the EFOSC2 spectra obtained one, two and six month after the super--outburst and measured the effective temperatures of the white dwarf cooling towards quiescence (Table~\ref{table:cooling}). 
The best-fit model to the spectrum taken one month after the super--outburst yields an effective temperature of $17\,000 \pm 2000\,$K. The strong continuum contribution arising from the disc and the contamination from the emission lines in the 2004--03--15 spectrum prevent a more accurate measurement of the white dwarf effective temperature.

For the 2004--08--27 we measured a white dwarf effective temperature of $\simeq 11\,700\,$K (Figure~\ref{lib_wd_2}). 
From the best fit model we found that the isothermal hydrogen slab is optically thin with a temperature of about $6200\,$K. These values are consistent with the accretion disc model of quiescent dwarf novae \citep{Williams1980,Tylenda1981}.

\subsection{The quiescence spectra}
The X--shooter average spectrum of QZ\,Lib, taken about ten years after its super--outburst is shown in Figure~\ref{xshooter_ave}. We compared the X--shooter spectra from the different years, i.e. 2012, 2013, and 2015 but found no difference and hence combined them to increase the SNR. This late spectrum differs from the spectra obtained within the first year of the outburst, showing stronger Balmer and helium emission lines, typical of a dwarf nova disc in quiescence (Table~\ref{tab:lines}).

We inspected the red part of the spectrum for absorption features arising from the secondary photosphere. Typically, when the donor star dominates the near--infrared emission, the most prominent absorption lines are \ion{Na}{i} $11\,381/11\,403\,$\AA, \ion{K}{i} $11\,690/11\,769\,$\AA\, and $12\,432/12\,522\,$\AA. However we could not identify any of those and therefore we assumed that the companion star contribution is negligible. We performed a spectral fit to the X--shooter data including a white dwarf and an hydrogen slab in the model. The best--fit model is shown in the top panel of Figure~\ref{xshooter_fit_sec} and returns a white dwarf effective temperature of $T_\mathrm{eff} = 10\,500 \pm 1500\,\mathrm{K}$, in agreement with the value determined by \citeauthor{Pala} (\citeyear{Pala}, $T_\mathrm{eff} = 11\,303 \pm 238\,\mathrm{K}$) from the analysis of ultraviolet data. The best--fit parameters for the hydrogen slab are listed in Table~\ref{table:slab_par} while the overall system parameters derived from the analysis of the EFOSC2 and X--shooter data are listed in Table~\ref{table:sys_par}. 

Although a clear signature of the companion star cannot be identified in the X--shooter data, only accounting for the white dwarf and the slab emission does not allow to adequately reproduce the observed flux level for $\lambda \gtrsim 10\,000\,$\AA. We therefore subtracted the white dwarf and the slab model to the X--shooter data and fit the grid of low--mass main sequence star (see Section~\ref{sec:model}) to the residual spectrum, constraining the secondary to be at the same distance as the white dwarf. The best--fit model is shown in the bottom panel of Figure~\ref{xshooter_fit_sec} and returns a secondary star effective temperature of $T_\mathrm{eff} \simeq 1700\,\mathrm{K}$. A discussion on the interpretation of this infrared excess is presented in the next Section.

\begin{table}
\caption{Best fit parameters for the isothermal and isobaric hydrogen slab describing the emission from accretion disc in QZ\,Lib.}\label{table:slab_par}
\begin{center}
\begin{tabular}{lcc}  
\toprule
Hydrogen slab parameter & Value\\
\midrule
Effective temperature (K) & $7100 \pm 300$ \\
Pressure (dyn\,cm$^{-2}$) & $110 \pm 40$ \\
Radial velocity (km\,{s}$^{-1}$) & $1150 \pm 250$\\
Inclination (\degree) & $30 \pm 12$\\
Geometrical height (cm) & $1.2(2) \times 10^7$\\
\bottomrule
\end{tabular}
\end{center}
\end{table}

\begin{figure*}
\centering
\includegraphics[angle=-90,width=\textwidth]{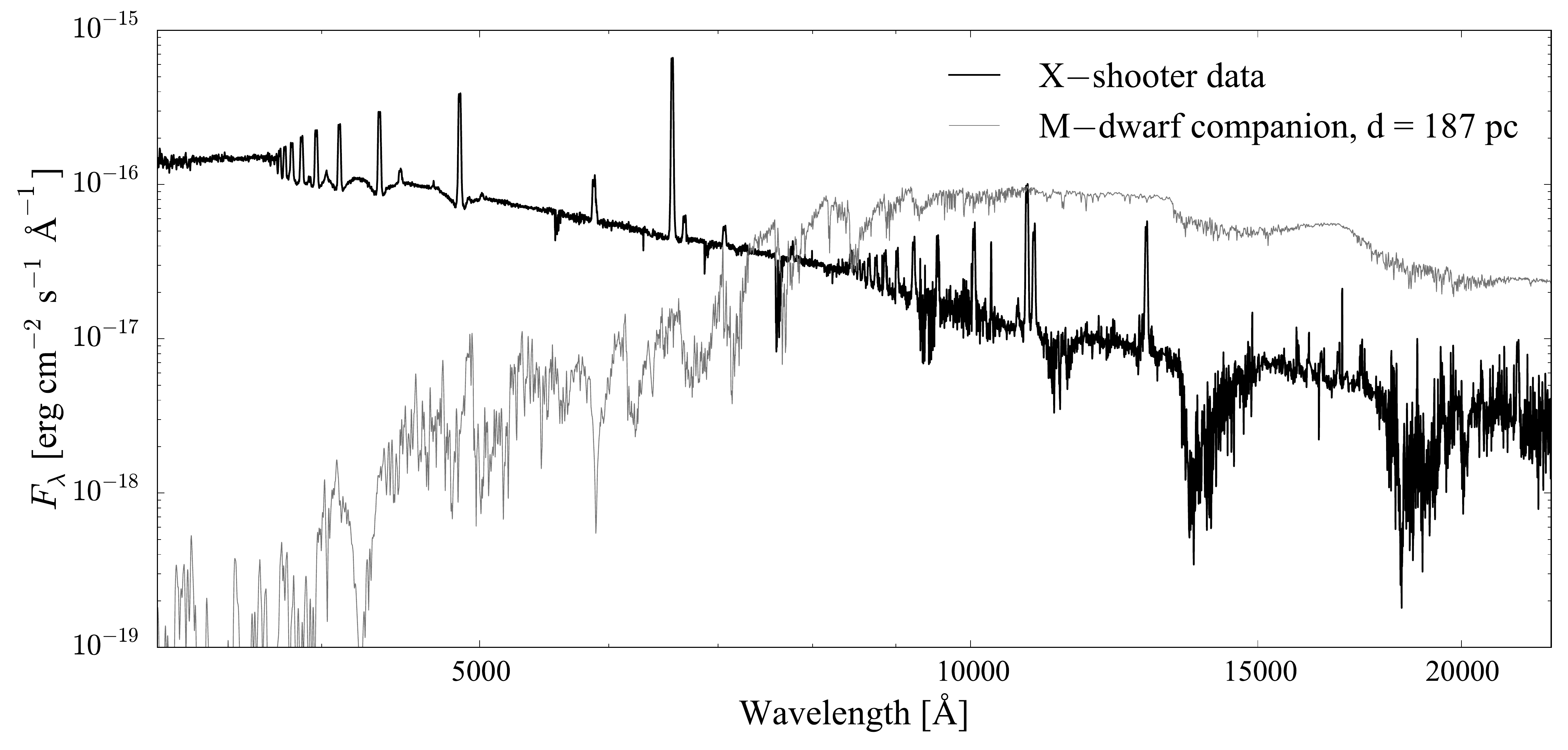}
\caption{X--shooter spectrum of QZ\,Lib (black) in comparison with the expected emission of a M6.5--dwarf companion as typical for a pre--bounce system with $P_\mathrm{orb} \simeq 90\,\mathrm{min}$ at a distance $d = 187\,$pc (grey) as derived from the \textit{Gaia} parallax.}\label{xshooter_mdwarf} 
\end{figure*}

\begin{figure*}
\includegraphics[width=\textwidth]{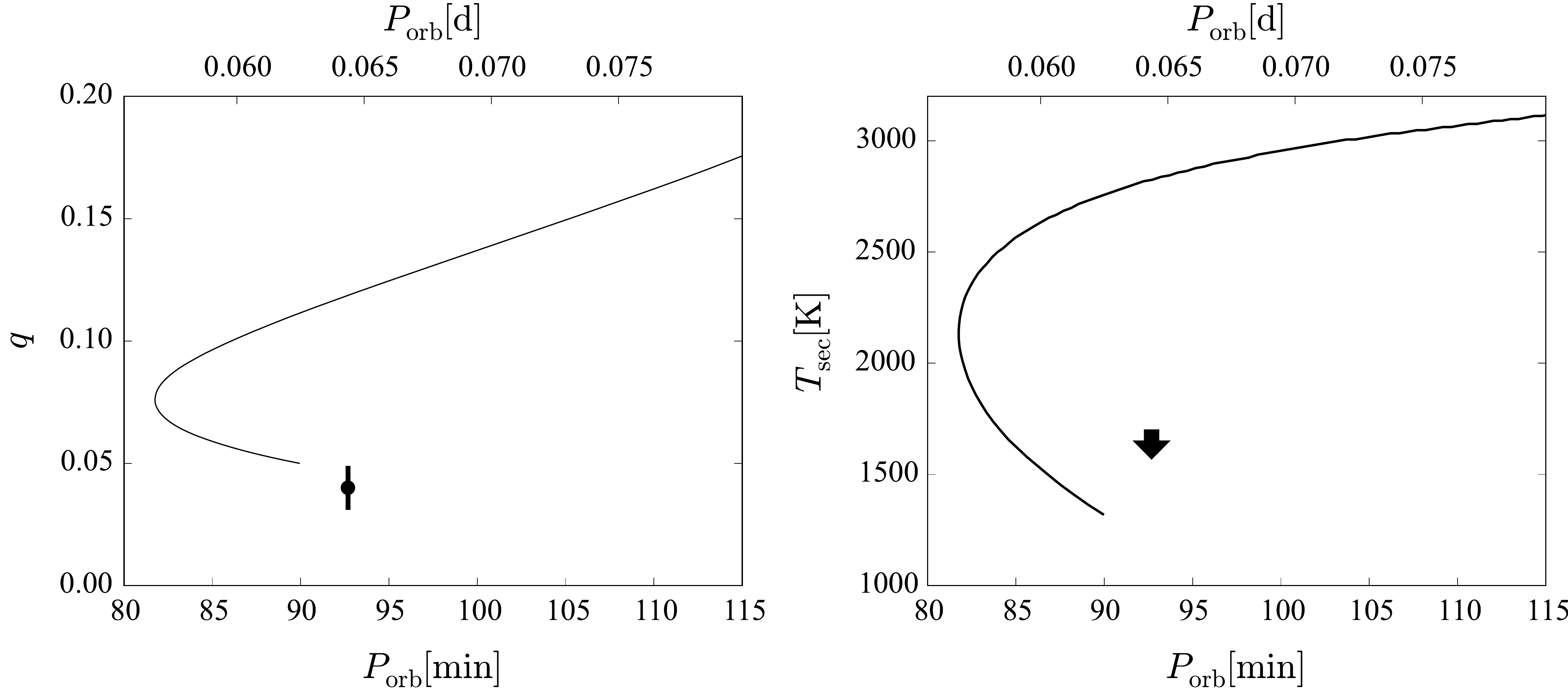}
\caption{QZ\,Lib mass ratio (left) and the upper limit on the effective temperature of its donor star (right) in comparison with the CV evolutionary track from \citeauthor{kniggeetal11-1} (\citeyear{kniggeetal11-1}, solid line, calculated assuming $M_\mathrm{WD} = 0.8\,\mathrm{M}_\odot$ for the left panel). These measurements clearly suggests a period bouncer system rather than a pre--bounce CV.}\label{q_t2_porb}
\end{figure*}

\begin{table}
\caption{System binary parameters for QZ\,Lib.}\label{table:sys_par}
\begin{center}
\begin{tabular}{lcc}  
\toprule
System parameter & Value & Instrument/Reference\\
\midrule
$P_\mathrm{orb}$ (min)                                 & $92.7 \pm 0.3 $          & EFOSC2    \\
$q$                                                    & $0.040 \pm 0.009$        & EFOSC2 \\
$d$ (pc)                                               & $187 \pm 12$             & \textit{Gaia}    \\ 
$i\,$(\degree)                                          &  $30 \pm 12$   &   X--shooter \\
$T_\mathrm{eff}$ (K)                                   & $10\,500 \pm 1500$     & X--shooter\\ 
$T_\mathrm{sec}$ (K)                                   & $\lesssim 1700$          & X--shooter\\
$\gamma\,(\mathrm{km\,s}^{-1}$)             & $-1.6 \pm 1.5$           & EFOSC2    \\
$K_\mathrm{WD}\,(\mathrm{km\,s}^{-1}$)                  & $20 \pm 4$               &  X--shooter  \\
$\log g$                                                                   & 8.35   &    fixed \\
$M_\mathrm{WD}\,(\mathrm{M}_\odot)$                 &  0.8    &    fixed       \\
$R_\mathrm{WD}\,(\mathrm{R}_\odot)$                  &  0.01   &   fixed \\
$M_\mathrm{sec}\,(\mathrm{M}_\odot)$                 &  0.032    &    fixed       \\
$R_\mathrm{sec}\,(\mathrm{R}_\odot)$                  &  0.10  &   fixed \\
\bottomrule
\end{tabular}
\end{center}
\end{table}

\section{Discussion}
We used VLT/X--shooter optical and near--infrared spectroscopy to reconstruct the Spectral Energy Distribution (SED) of QZ\,Lib, from $\lambda \simeq 3000\,$\AA\, out to $\lambda \simeq 23\,000\,$\AA. From a spectral fit to the data, we have shown that the contribution from a white dwarf and an isothermal slab cannot adequately reproduce the flux in the red portion of the spectrum. This infrared excess can be explained by either the accretion disc and/or a brown dwarf secondary star. 

The isothermal and isobaric hydrogen slab we included in the fit procedure is a good model to account for the emission lines that arise from the optically thin upper layers of the disc. However, if the central and colder regions of the disc are composed of an optically thick gas, they could contribute to the overall SED of the system in the form of an additional blackbody(--like) emission that is not accounted for by our model. At present, not much is known about the spectrum arising from the cool portion of quiescent discs. In the past, this contribution has been modelled by dividing the disc into several annuli and then summing the emission of each annulus, approximated as a blackbody with an effective temperature set accordingly to the distance of the annulus from the white dwarf \citep{howell2006,Brinkworth2007,howell2008}. Moreover, since this gas has a similar composition and effective temperature as the secondary, is very difficult to unambiguously identify the near--infrared flux as either the disc or donor emission. Given that the wavelength coverage of X--shooter is limited at the $K$--band, additional constraints on the SED at longer wavelengths are needed to distinguish between a stellar component or a disc contribution. Infrared photometric observations of QZ\,Lib are available from the \textit{WISE} (Wide--field Infrared Survey Explorer) satellite. However, these photometric measurements are contaminated by the presence of a close background object and cannot provide any additional constraints on the origin of the observed infrared excess. In any case, a near--infrared excess arising exclusively from the disc would only imply a lower effective temperature of the secondary, bringing our results into an even better agreement with the predicted evolutionary track of the period bouncer CVs (see below).

Although we cannot distinguish between the emission of the donor star and that of a possibly optically thick cool accretion disc, given the mass ratio of QZ\,Lib, $q = 0.040(9)$, we can still conclude that its donor star must be a brown dwarf. In fact, considering a white dwarf mass $0.6\,\mathrm{M}_\odot \lesssim M_\mathrm{WD} \lesssim 1.44\,\mathrm{M}_\odot$, the secondary star mass results $0.024\,\mathrm{M}_\odot \lesssim M_\mathrm{sec} \lesssim 0.058\,\mathrm{M}_\odot$, i.e. below the hydrogen--burning limit and clearly in the brown dwarf regime. 
Furthermore the expected spectral type of the donor star in a pre--bounce CV at $P_\mathrm{orb} \simeq 90\,\mathrm{min}$ is M6.5, corresponding to $M_\mathrm{sec}\,=\,0.096\,\mathrm{M}_\odot$, $R_\mathrm{sec} = 0.144\,\mathrm{R}_\odot$ and $T_\mathrm{eff} = 2744\,\mathrm{K}$ \citep{Knigge2006}, is not consistent with the observed SED. In fact, such a star would clearly dominate the spectral appearance for $\lambda \gtrsim 7000\,$\AA\, (Figure~\ref{xshooter_mdwarf}). Instead, our fit to the X--shooter data shows that the upper limit on the secondary effective temperature is $T_\mathrm{eff} \simeq 1700\,\mathrm{K}$, supporting the conclusion of a brown dwarf donor star (Figure~\ref{q_t2_porb}, right panel). 

As discussed above, the mass of the secondary star, i.e. the system mass ratio, is one of the strongest discriminants between pre and post--bounce systems (see e.g. figure~6 from \citealt{howelletal01-1}).
Inspecting the location of QZ\,Lib in the $P_\mathrm{orb}-q$ distribution shown in the left panel of Figure~\ref{q_t2_porb} strongly suggests that it is in the regime of potential period bouncers, rather than on the pre--bounce evolutionary track.
Owing to its low mass ratio QZ\,Lib has been previously suggested as a potential period bouncer by \cite{Patterson2011} and, subsequently, also by \cite{Pala} because of its cool white dwarf. Our fit to the SED of QZ\,Lib provides additional support in favour of the period--bouncer nature of QZ\,Lib. 
 
Alternatively, QZ\,Lib could have been born with a brown dwarf donor star. These kind of binaries start mass transfer below the period minimum and then evolve towards longer orbital periods, finally merging into the standard track of period bouncers (figure~1 from \citealt{kolb+baraffe99-1}). Even if this scenario cannot be ruled out, the population synthesis study from \citet{Politano2004} has shown that CVs hosting brown dwarf donors with orbital periods above the period minimum are four times more likely to have followed the standard path of CV evolution (i.e. the donor has become substellar as a consequence of mass erosion during the CV evolution) than being born as a white dwarf plus brown dwarf binary, suggesting that QZ\,Lib is more likely a period bouncer CV.

\section{Conclusions}
We present photometric and spectroscopic time--resolved observations of the cataclysmic variable QZ\,Lib. From radial velocity measurements of the H$\alpha$ emission line, we determine the orbital period $P_\mathrm{orb} = 0.06436(20)\,$d that, combined with the superhump period, $P_\mathrm{SH} = 0.06557(14)$\,d, yields the system mass ratio, $q = 0.040 \pm 0.009$. Assuming the mass of the white dwarf to be in the range $0.6\,\mathrm{M}_\odot \lesssim M_\mathrm{WD} \lesssim 1.44\,\mathrm{M}_\odot$, we found that the secondary mass is $0.024\,\mathrm{M}_\odot \lesssim M_\mathrm{sec} \lesssim 0.058\,\mathrm{M}_\odot$, and therefore the donor must be a brown dwarf.

From a spectral fit to the averaged X--shooter spectrum we measure the white dwarf effective temperature, $T_\mathrm{eff} = 10\,500 \pm 1500\,\mathrm{K}$. Modelling the data with a white dwarf and an isothermal and isobaric hydrogen slab does not adequately reproduce the observed flux in the reddest portion of the spectrum. Interpreting this weak infrared excess as the contribution from the donor star implies an upper limit on its effective temperature, $T_\mathrm{sec} \lesssim 1700\,\mathrm{K}$. However, this excess could also originate from an optically thick accretion disc. The limited wavelength range of our observations does not allow to further investigate this possibility but, if the near--infrared excess arises from the disc, it would imply the presence of an even colder secondary star, bringing our results into an even better agreement with the predicted evolutionary track of the period bouncer CVs.

Although we cannot exclude that QZ\,Lib is born with a brown dwarf donor, this scenario would be four times less likely than the detection of a period bouncer CV. We therefore argue that QZ\,Lib is a CV that has evolved through the period minimum. In fact, QZ\,Lib meets all the requirements for being a period bouncer: it has a brown dwarf companion, a low mass ratio and hosts the coolest white dwarf at the longest orbital period (see its position in figure~21 from \citealt{Pala}). This is the first CV for which all these requirements have been spectroscopically confirmed, ruling out a pre--bounce system still on its way to the period minimum and, instead, making QZ\,Lib the strongest period bouncer candidate identified so far. 

The detection of period bouncer CVs is an important test for the theoretical prediction of the standard model of CV evolution, accordingly to which $\simeq 40-70\%$ \citep{kolb93-1,kniggeetal11-1,Goliasch-Nelson2015} of the present Galactic CV population is expected to have evolved past the period minimum. However, despite more than 40 years of intense research on CVs which has led to the identification of over 1400 systems with an orbital period determination \citep{ritter+kolb03-1}, only a handful of period bouncer have so far been identified. QZ\,Lib illustrates how some of these elusive systems could be hiding among the known CVs but have not been identified as such owing to the limited wavelength range in which they have been studied so far. We illustrated how multi--wavelength observations, extending from the ultraviolet to the infrared, are one of the most powerful tool to unveil this major component of the present day Galactic CV population.

\section*{Acknowledgements}
Based on observations made with ESO Telescopes at La Silla and Paranal Observatories under programme ID 60.A-9013(A), 089.D-0421(A), 095.D-0802(A).\\
This work has made use of data from the European Space Agency (ESA) mission {\it Gaia} (\url{https://www.cosmos.esa.int/gaia}), processed by the {\it Gaia} Data Processing and Analysis Consortium (DPAC, \url{https://www.cosmos.esa.int/web/gaia/dpac/consortium}). Funding for the DPAC has been provided by national institutions, in particular the institutions participating in the {\it Gaia} Multilateral Agreement.\\
The research leading to these results has received funding from the European Research Council under the European Union's Seventh Framework Programme (FP/2007--2013) / ERC Grant Agreement n. 320964 (WDTracer). \\
A.~F.~P. gratefully acknowledges the support of the Visiting Scientists Program at ESO for making possible her stays in Santiago, during which part of this work was done. C.~T. acknowledges support by the Centro de Astrof\'isica de Valpara\'iso (CAV).\\
We acknowledge with thanks the variable star observations from the AAVSO International Database contributed by observers worldwide and used in this research.

\bibliographystyle{mn2e}
\bibliography{aamnem99,aabib}

\bsp

\label{lastpage}

\end{document}